\documentclass[a4paper, 11pt]{article}

\usepackage{fullpage}
\usepackage{amsmath,amsthm}
\usepackage{amssymb, bbm}
\usepackage{epsfig}
\usepackage{stmaryrd}
\usepackage[usenames]{color}
\usepackage[hang,flushmargin]{footmisc}
\usepackage{graphicx}
\usepackage{dsfont}
\usepackage{titlesec}
\usepackage[affil-it]{authblk}
\usepackage{setspace}
\usepackage{titling}
\usepackage{graphicx,enumerate,mathrsfs,color,subfigure}
\usepackage{mathtools}
\usepackage{sidecap}

\titleformat{\subsection}[runin]{\normalfont\bfseries}{\thesubsection.}{3pt}{}[.]
\titleformat{\section}[runin]{\large\bfseries}{\thesection.}{1pt}{}[]

\def\det{\operatorname{det}}

\def\Curl{\operatorname{Curl}}

\numberwithin{equation}{section}

\def\dsp{\def\baselinestretch{1.3}\large\normalsize}
\dsp

\newcommand\mbf[1]{\ensuremath{\mbox{\boldmath$#1$}}}

\theoremstyle{remark}
\newtheorem{rem}{Remark}[section]

\begin{document}

\title{Geometry of Defects in Solids}

\author{Ayan Roychowdhury and Anurag Gupta\thanks{ag@iitk.ac.in}}
\date{{\small Department of Mechanical Engineering, Indian Institute of Technology, Kanpur 208016, India
\\  \today}}
\maketitle

\begin{quote}
\singlespacing
\footnotesize{\ldots {\it true geometry} [is] a doctrine of {\it space itself} and not merely like Euclid, and almost everything else that has been done under the name of geometry, a doctrine of the configurations that are possible in space.

\vspace{-5pt}

\hspace{30pt} Hermann Weyl\footnote{p. 102 in H. Weyl, {\it Space-Time-Matter} (translated by H. L. Brose), Dover, 1952.} }
%\footnotesize{A science can only determine its domain of investigation up to an isomorphic mapping. In particular it remains quite indifferent as to the `essence' of its objects. That which distinguishes the real points in space from number triads or other interpretations of geometry one can only {\it know} by immediate intuitive perception....The idea of isomorphism demarcates the self-evident insurmountable boundary of cognition. - Hermann Weyl\footnote{pp. 25-26 in {\it Philosophy of Mathematics and Natural Science}, Atheneum, 1963.} }
\end{quote}

\section{Introduction}

Up until early nineteenth century, the notion of geometry was unambiguously Euclidean. Based on fundamental objects such as points, straight lines, and planes, and a set of elementary presuppositions (axioms) about their mutual relationship, the predominant focus in Euclidean geometry is to derive, in a logically consistent manner, ``the configurations that are possible in space".\footnote{This basis for geometry remained ``self-evident" for more than twenty-two centuries. An occasional discomfort was caused by the parallel postulate of Euclidean geometry, but never to an extent of questioning the validity of the geometry.} The ``space itself" remains {\it continuous} (i.e. between any two points in space there are infinitely many points), {\it locally flat} (i.e. the solid angle around any point in space is same; it is $2\pi$ for a two-dimensional Euclidean plane and $4\pi$ for a three-dimensional Euclidean space), {\it homogeneous} (i.e. a body can move in space without changing its size and shape), and {\it similar} (i.e. a body can be reconstructed to any scale in another part of space).\footnote{The concept of space was absent from Euclidean geometry until the insightful work of Descartes. The ancient geometers took the nature of space for granted and busied themselves solely with understanding the character of geometrical figures which could occupy it. The four postulates about the nature of space that are mentioned here were given by Clifford, see pp. 210-230 in W. K. Clifford, {\it Lectures and Essays}, Macmillan and Co., 1886. In this work of great originality Clifford demonstrated the equivalence of these postulates with the axioms of Euclidean geometry.} That such is the true nature of space was an unshakeable belief held alike by philosophers, mathematicians, and physicists. In particular, no scientific theory of the physical world was expected to be in discord with this structure of the space and hence with the propositions of Euclidean geometry.

The pioneering non-Euclidean revolution was brought about by the Russian mathematician Nikolai Invanovich Lobachevski---``what Copernicus was to Ptolemy, that was Lobachevski to Euclid"\footnote{Clifford, {\it op. cit.}, p. 212.}---who constructed a geometry by providing an alternative to Euclid's parallel postulate. This in effect introduced the possibility of a geometrical space which is continuous, locally flat, and homogeneous, but not similar; such spaces necessarily have a constant non-zero curvature (Euclidean space has zero curvature). Lobachevski's construction led to spaces with constant negative curvature; the other possibility, of spaces with constant positive curvature, was suggested several decades later by Riemann. Most importantly, Lobachevski's contribution exposed our fallibility of considering Euclidean geometry as the irreplaceable truth of nature.

The second breakthrough in non-Euclidean geometries came from the German mathematician Georg Friedrich Bernhard Riemann. Inspired by Gauss's theory of surfaces, Riemann considered spaces (of arbitrary dimension) which are continuous and locally flat, but not necessarily homogeneous and similar. He characterized them in terms of a {\it metric} field which generates a quadratic form representing the distance between infinitesimally closed points. For Riemannian spaces, knowing the metric function is sufficient to determine the parallel transport of vectors and the curvature of space; the latter no longer remaining a scalar constant. The ingenuity of Riemann was to interpret metric not as an {\it a priori} property of the space but instead as a characteristic of the ``physical phenomenon" manifested in an otherwise formless space. Hence unlike both Euclidean spaces and Gaussian surfaces, where the metrical properties are fixed once for all, the Riemannian metric, and consequently the resulting geometry, is allowed to be derived from the ``matter" filling the space and ``the binding forces which act upon it".\footnote{cf. B. Riemann, ``On the Hypotheses which lie at the Bases of Geometry" (1854), translated by W. K. Clifford, {\it Nature}, Vol. VIII. Nos. 183, 184, pp. 14--17, 36, 37, 1873.} The geometry, rather than acting merely like a rigid skeleton in the background of a physical theory, was now free to participate in it.\footnote{Aptly summarized by Weyl (p. 220., {\it op. cit.}), ``this seals the doom of the idea that a geometry may exist independently of physics".} The physical relevance of Riemannian geometry remained completely unappreciated for as long as seventy years, until after the appearance of Einstein's theory of general relativity wherein the metrical structure of the four-dimensional space-time continuum was identified with the gravitational field associated with the ``matter" occupying the continuum. Fortunately Riemann's theory met with an all together different fate in the hand of mathematicians; by the time that Einstein's theory made its appearance, it had already achieved maturity in the well established disciplines of tensor analysis and differential geometry.

%The developments in seemingly disjoint disciplines of scientific inquiry have had, and continue to have, a much greater influence on each other than what meets even an informed eye. The theory of elasticity of solids, which was once the precursor to the later field theories of electromagnetism and relativity,\footnote{Baron Augustin-Louis Cauchy, a brilliant French mathematician, was the first one to construct a truly continuum theory while working towards a field theory of elasticity. His work paved the way for Maxwell's theory of electromagnetism leading to Einstein's theory of relativity, cf. Ch. IV in M. Born, {\it Natural Philosophy of Cause and Chance}, Dover, 1964.} has in turn benefitted significantly from them in the last century. The study of defects in a solid continuum serves as an illustrious example, where the concepts of force acting on a defect and motion of a dislocation were, for instance, introduced as analogous to the force acting on a point charge and the motion of particles in special relativity, respectively.\footnote{cf. p. 62 and p. 107 in X. Markenscoff and A. Gupta (Eds.), {\it Collected works of J. D. Eshelby}, Springer, 2006.}

The success of non-Euclidean geometries with the relativity theory provided impetus to their application in other domains of mechanics including development of a elasticity theory of a continuous distribution of defects.\footnote{The theory of elasticity of solids, which acted as the precursor to the later field theories of electromagnetism and relativity, has in turn benefitted significantly from them in the last century. Besides motivating a non-Euclidean framework, there are several other instances where these later field theories have assisted elasticity with fundamental breakthroughs. The force acting on an isolated defect in a solid body and the motion of a dislocation were, for instance, introduced as analogous to the force acting on a point charge and the equation of motion of particles in special relativity, respectively,   cf. p. 62 and p. 107 in X. Markenscoff and A. Gupta (Eds.), {\it Collected works of J. D. Eshelby}, Springer, 2006.} Whereas it is the geometry of the four-dimensional space-time which is treated as non-Euclidean in a relativistic continua, it is the geometry of the material space which is most naturally described as non-Euclidean in a defective continua. Interestingly, while it is the ``matter" which induces curvature in a relativistic space-time continuum, it is the presence of defects (apparent lack of ``matter") which brings about a curvature in the material space. The material space can be thought of as a natural configuration of the body where it is described only in terms of the intrinsic structure of the constituting matter. The material space of a crystalline solid, for instance, is the configuration obtained by relaxing the solid of all internal and external stresses. The relaxed configuration of a defective solid will not be a coherent body in the Euclidean space. The geometric experience of the imaginary beings living in the material space would be Euclidean in the absence of any defects, but non-Euclidean when the body is defective. These beings, otherwise insensitive to distortions caused by any external agency (such as load, temperature field, etc.), will recognize geometrical evolution of their space of existence only with appearance or disappearance of defects.\footnote{cf. p. 287 in E. Kr\"{o}ner, Continuum theory of defects, {\it Les Houches, Session XXXV, 1980 -- Physique des d{\'e}fauts}, North-Holland, 1981 and Ch. IV in H. Poincar\'{e}, {\it Science and Hypothesis}, The Walter Scott Publishing Co., Ltd., 1905.} To bring forth the appealing connection between non-Euclidean geometries and defects in solids is the purpose of this article.

Drawing a correspondence between the nature of a defect and a specific geometric property of the material space not only illuminates the underlying structure of defects in solids but also provides an unambiguous way to represent defect densities within a physical theory. The defect characterization can be used to calculate the internal stress field in an elastic solid, or to represent hardening during plastic deformation.\footnote{cf. Kr\"{o}ner, {\it op. cit.}, and F. N. R. Nabarro and M. S. Duesbery (Eds.), {\it Dislocations in Solids V. 11}, North-Holland, 2002.} It also serves as a macroscopic representative for the microstructure and is a natural device to introduce microscopic length scales in the theory.\footnote{Nabarro and Duesbery, {\it op. cit.}} Most importantly the kinetic laws for the dynamic evolution of the material body are conveniently expressed as partial differential equations for defect densities.\footnote{\it Ibid.} A geometrical study of defects can be motivated from these, among several other practical applications, if not only from the sheer elegance of its mathematical structure.

To accommodate a broad spectrum of readership we have kept the discussion sufficiently self-contained and accessible. References are provided generously with an intention to lead the interested reader to further directions as well as background information. We have divided the rest of this article into three parts. The first part, given in Section \ref{geom} below, deals with concepts from differential geometry. Beginning with a quick overview of non-Euclidean geometries, we move on to a brief introduction of tensor analysis and differential geometry while restricting ourselves mostly to the concepts useful for our present discussion. In Section \ref{lattice}, isolated defects in a two-dimensional lattice are considered and their non-Euclidean nature is emphasized. We observe that the presence of a dislocation in a lattice leads to the failure of parallelogram closure, of a disclination to the failure of distant parallelism, and of a point defect to the non-uniformity of the metric.\footnote{Parallelogram closure, distant parallelism, and uniformity of metric are all essential features of a Euclidean space.} In the last section of this article, these observations are used to motivate the relationship between continuous defect densities and various tensor fields in differential geometry.  In particular, we identify the Riemann-Christoffel tensor (or curvature tensor), the Cartan tensor (or torsion tensor), and the nonmetricity tensor (obtained from the covariant derivative of the metric tensor), associated with the material space, with the density of disclinations, dislocations, and point-defects (vacancies, interstitials, substitutional), respectively. We end our discussion with remarks on the elastic stress field associated with defect distribution and on the analogy between the present theory and the general theory of relativity.

\section{Geometry}

The origin of geometry, according to a viewpoint, is in the ritualistic traditions of ancient civilizations.\footnote{Sumerians and Babylonians used geometrical drawings in constructing temples and performing augury, respectively. The Egyptian priests knew of similar triangles and how to calculate the volume of a truncated pyramid while the Vedic priests knew enough geometry, as demonstrated in their manuals of altar construction ({\it \'{S}ulvas\={u}tras}), to attempt converting a rectangle (or a circle) into a square and a square into a circle (of equal area). These highly original manuals, which predate Greek mathematics by at least a few centuries, also contain a geometrical version of the Pythagoras' theorem, among several other interesting results. For further reading on {\it \'{S}ulvas\={u}tras} and ritual origins of geometry see A. Seidenberg, {\it Arch. Hist. Ex. Sci.}, 1, pp. 487-527, 1962 (also 18, pp. 301-342, 1978) and F. Staal, {\it J. Ind. Phil.}, 27, pp. 105-127, 1999.} It serves us well, however, to begin with Euclid's {\it Elements} (ca. 300 B.C.). Within Euclidean geometry a set of postulates (axioms), taken to be fundamental and undoubtable truths about the (geometric) nature of space, are used as the basis to derive a large number of geometrical theorems. The postulates, as well as the resulting theorems, confirm naturally to our everyday perception of the world; hence reinforcing the belief that Euclidean geometry is the true geometry for all physical phenomena. It was not until early nineteenth century when Lobachevski constructed an alternative geometry,\footnote{Also suggested independently by Gauss, Schweikart, and Bolyai.} called the hyperbolic geometry, by replacing Euclid's parallel postulate with the postulate that more than one straight line can be drawn from a point outside a given straight line in a plane, without intersecting the given line; all the other fundamental postulates of Euclidean geometry were retained.\footnote{According to the parallel postulate of Euclid, only one straight line passes through a given point, away from a given straight line in a plane, that does not intersect the given line. It should be noted that the notion of ``points" and ``straight lines" is abstract in the axiomatic structure of these geometries. They are to be given a physical interpretation while using a particular geometry to describe the space which satisfies the relevant postulates. One of the most accessible, yet masterly, treatment of axiomatic formulations of Non-Euclidean geometries can be found in Ch. IV of D. Hilbert and S. Cohn-Vossen, {\it Geometry and Imagination}, AMS Chelsea Publishing, 1999.} Thereafter, Riemann suggested another geometry, now known as the elliptic geometry, by instead assuming that no straight line can be drawn through the given point which will not intersect the given line. The geometry of Riemann require straight lines to be unbounded but finite, unlike Euclidean geometry where they are postulated to be both unbounded and infinite. It was later shown by Felix Klein that these non-Euclidean models of geometry are as consistent as the Euclidean geometry itself and hence equally suitable for studying a physical experiment, a prerogative hitherto ascribed only to the latter. In fact, in the words of Hermann Weyl, ``...the validity or non-validity of Euclidean geometry can not be proved by empirical observations....it is only the whole composed of geometry and physics that may be tested empirically".\footnote{p. 93 in Weyl, {\it op. cit.}}

\label{geom}

\subsection{Theory of Surfaces} The axiomatic formulations of geometry are mainly concerned with the global structure of space; their results (theorems, etc.) are derived in a purely constructive manner. Alternatively, geometry can be studied using tools from analysis and differential calculus. This is the realm of differential geometry; here the emphasis is to study the local structure of space, i.e. only within small neighborhoods of a point, while employing analytical methods in mathematics. The general theory of curved surfaces, as developed by Gauss in the early nineteenth century, forbears the development of this important discipline. We pause, only briefly, to highlight those features of this important theory which are directly relevant to our discussion.\footnote{There are many excellent expositions which can be used to study the theory of surfaces, see e.g. T. J. Willmore, {\it An Introduction to Differential Geometry}, Oxford University Press, 1959.} The two-dimensional surface, parameterized by $(u^1, u^2) \in \mathbb{R}^2$ ($\mathbb{R}$ denotes the set of real numbers), is assumed to be embedded in a three-dimensional Euclidean space with coordinates $(x^1, x^2, x^3) \in \mathbb{R}^3$ such that, for points on the surface, the coordinates can be expressed as sufficiently smooth functions of the surface parameters. The distance $ds$ between two infinitely near points on the surface is given in terms of a quadratic form, expressible either as\footnote{Here, and in rest of the article, we use Einstein's summation convention for repeated indices. The indices represented by lowercase Roman alphabets vary from $1$ to $3$ and those by lowercase Green alphabets vary from $1$ to $2$.}
\begin{equation}
ds^2 = dx^i dx^i \label{geom1}
\end{equation}
(i.e. a Pythagoras form) or equivalently, in terms of the surface parameters, as
\begin{equation}
ds^2 = h_{\alpha \beta} du^\alpha du^\beta, \label{geom2}
\end{equation}
where the {\it metric} functions
\begin{equation}
h_{\alpha \beta} = \frac{\partial x^k}{\partial u^\alpha}\frac{\partial x^k}{\partial u^\beta} \label{geom3}
\end{equation}
are clearly dependent on the two surface parameters alone; no generality is lost in assuming $h_{\alpha \beta} = h_{\beta \alpha}$. It is important to note that the surface has a two-dimensional Euclidean nature in a small neighborhood of every point; i.e. there always exist a (local) re-parametrization of the surface with respect to which the infinitesimal distance \eqref{geom2} is reduced to a Pythagoras form with $h_{11} = h_{22} = 1$ and $h_{12}=h_{21}=0$. This change in the parametrization will, in general, be different for different points on the surface. If however there exists a single re-parametrization which transforms the distance \eqref{geom2} at all points on the surface to a Pythagoras form then the surface is {\it flat}.

Knowing the metric is sufficient to determine the length of curves, the angle between curves, and the area enclosed by closed curves on the surface. Such measurements which can be made by a geometer who is completely confined to the surface, oblivious to the three-dimensional space surrounding the surface, are called intrinsic. All intrinsic characteristics of the surface are derivable from the metric. Two important intrinsic features of the surface are given by the {\it connection} functions $\Gamma^{\alpha}_{\gamma \beta}$ (eight in number) and the {\it Gaussian curvature} $K$ (scalar function). The former is related to the (differentiable) metric functions as
\begin{equation}
\Gamma^{\alpha}_{\gamma \beta} = \frac{1}{2} h^{\delta \alpha} \left( \frac{\partial h_{\beta \delta}}{\partial u^\gamma} + \frac{\partial h_{\gamma \delta}}{\partial u^\beta} - \frac{\partial h_{\beta \gamma}}{\partial u^\delta} \right) \label{geom3.1}
\end{equation}
where $h^{\alpha \delta}$ are functions which satisfy $h_{\alpha \gamma} h^{\alpha \beta} = \delta_\gamma^\beta$.\footnote{$\delta_\gamma^\beta = 1$ if $\gamma = \beta$ but $0$ otherwise.} These functions are symmetric in the sense that $\Gamma^{\alpha}_{\beta \gamma} = \Gamma^{\alpha}_{\gamma \beta}$. The form of the Gaussian curvature in terms of the metric functions is voluminous and can be seen elsewhere.\footnote{cf. p. 79 in Willmore, {\it op. cit.}} We will revisit these features, in a more systematic way, for three-dimensional Riemannian spaces in the following.

\begin{figure}[t!]
\centering
\subfigure[]{\includegraphics[scale=0.30]{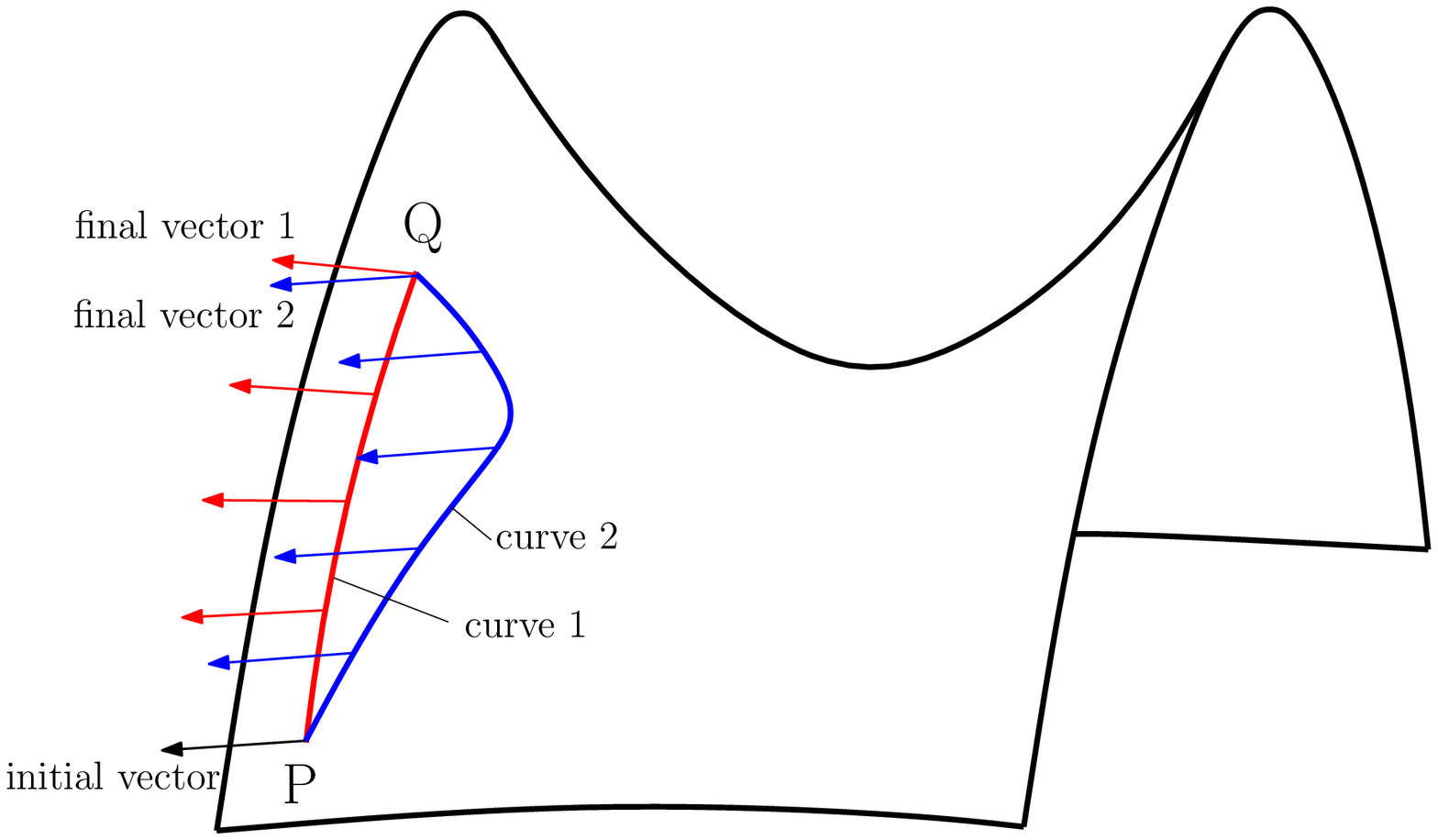}}
\hspace{10mm}
\subfigure[]{\includegraphics[scale=0.30]{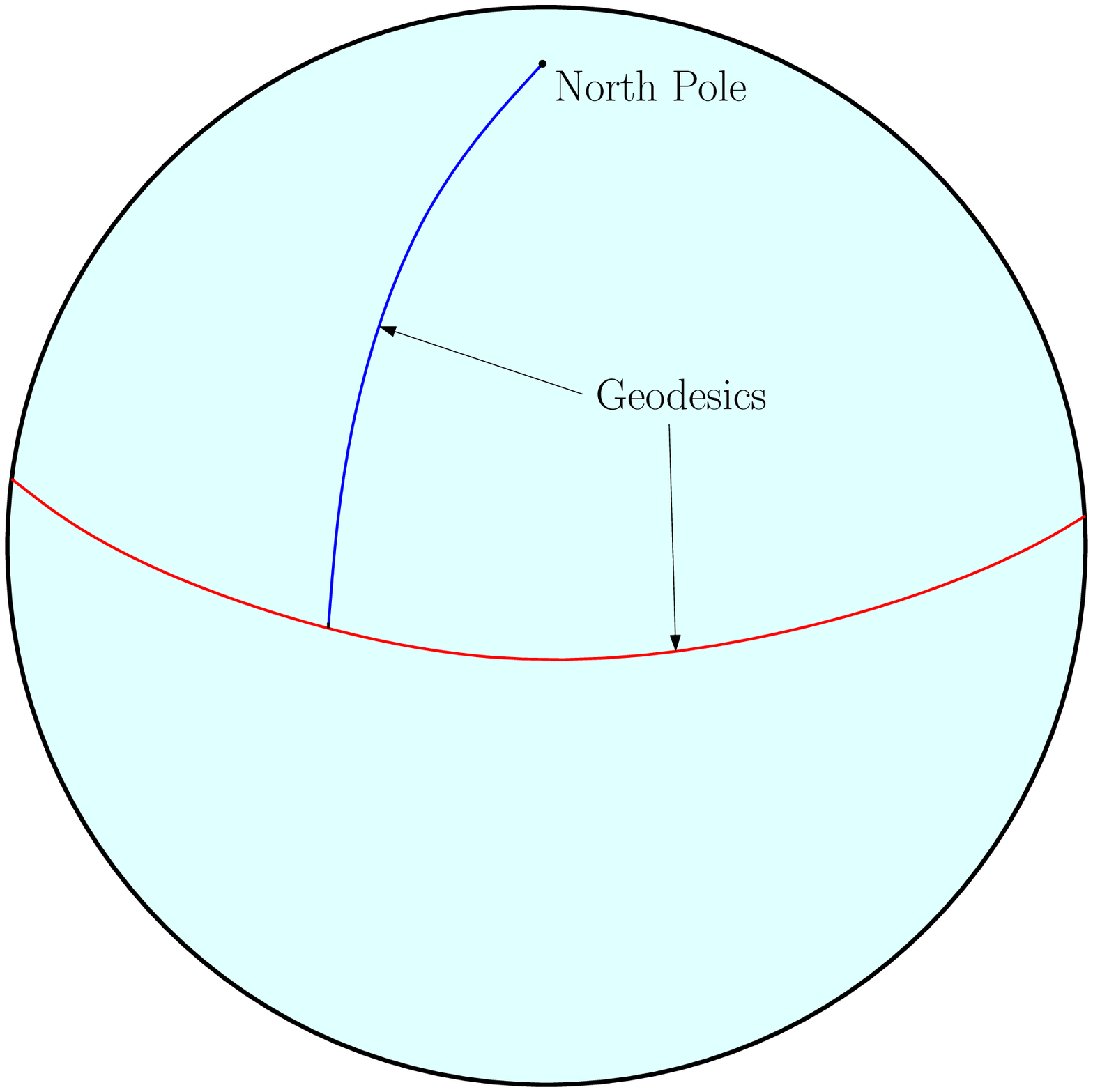}}
\caption{(a) Parallel transport of a vector between two points $P$ and $Q$ along two curves on a hyperbolic paraboloid. (b) Geodesics or great circles on a spherical surface can be interpreted as straight lines.}
\label{surfacesf}
\end{figure}

The connection determines how a vector is transported parallelly along a curve on the surface. {\it Parallel transport}, intuitively speaking, brings about the change in a vector from one tangent plane on the surface to another (when viewed in the embedding space) only as much as is {\it forced} by the curvature of the surface. The Gaussian curvature on the other hand determines whether the vector obtained by parallel transporting a given vector, between two fixed points on the surface, depends on the curve along which it is transported, see Figure \ref{surfacesf}(a). The Gaussian curvature is everywhere zero for a flat surface.

Interestingly, with suitable definition of points and straight lines, surfaces with constant positive and negative Gaussian curvature can be identified with the two-dimensional spaces of elliptic and hyperbolic geometry, respectively. For example, the geometry on a spherical surface, cf. Figure \ref{surfacesf}(b), can be regarded as an instance of the former case if we identify any great circle with a straight line and any two diametrically opposite points as one point.

\subsection{Affine connection on a three-dimensional differentiable manifold} Before proceeding to discuss Riemann's generalization of Gauss's theory, as well as some other pertinent results from differential geometry, we introduce some preliminaries from tensor analysis; in particular the notion of tangent space, affine connection (and the torsion and curvature tensors associated with it), and covariant derivative.\footnote{Our treatment of the subject is based on Willmore, {\it op. cit.} See also J. A. Schouten, {\it Ricci-Calculus}, Springer-Verlag, Berlin, 1954.} Let $\mathcal{X}$ be a three-dimensional differentiable manifold such that for each point of the manifold there exists a neighborhood which is homeomorphic to some open set in a three-dimensional Euclidean space.\footnote{cf. pp. 193-194 in Willmore, {\it op. cit.}, for the formal definition of a differentiable manifold. The term manifold should be understood to carry the same connotation as the term (topological) {\it space} used informally elsewhere.} Such neighborhoods are called coordinate neighborhoods. We can associate a local coordinate system over the coordinate neighborhood $U$ with respect to which every point in $U$ is represented by a triad of real numbers, denoted by $x^i$ (with $i= 1,2,3$). Let $\gamma$ be a curve on $\mathcal{X}$ whose intersection with $U$ is given by points $x^i$ written as differentiable functions $f^i(t)$, where $0 \leq t \leq 1$ is a parameter such that $f^i(0) = x^i_0$ represents a point $P$ in $U$. A {\it tangent vector} to $\gamma$ at $P$ is a vector whose components, with respect to the natural basis associated with the coordinate system, are given by $\dot{x}^i_0$ where
\begin{equation}
\dot{x}^i_0 = \frac{d f^i}{d t}\Big|_{t=0}. \label{geom4}
\end{equation}
It is straightforward to see that the set of tangent vectors to all curves passing through $P$ form a three-dimensional vector space, called the {\it tangent space} of $\mathcal{X}$ at $P$ and denoted by $\mathcal{T}_P$. We will often use, without any loss of generality, $dx^i$ to denote components of a tangent vector at $x^i$.

The tangent spaces at different points of the manifold are all three-dimensional, and hence isomorphic to each other.\footnote{Two vector spaces $\mathcal{U}$ and $\mathcal{V}$ are isomorphic if there is a one-to-one correspondence between the vectors ${\bf x} \in \mathcal{U}$ and the vectors ${\bf y} \in \mathcal{V}$, say ${\bf y} = T({\bf x})$, such that $T(\alpha{\bf x} + \beta{\bf y}) = \alpha T({\bf x}) + \beta T({\bf y})$; cf. P. R. Halmos, {\it Finite-Dimensional Vector Spaces}, Springer, 1987.} A {\it connection} is an additional structure on $\mathcal{X}$ which gives a definite isomorphism relating any two tangent spaces on the manifold. Consider a curve $\gamma$ on $\mathcal{X}$, as introduced above, and let $\mathcal{T}_t$ be the tangent space at the point on the curve given by parameter $t$; this point shares the same coordinate neighborhood $U$ with $P$. An isomorphism $\phi_t:\mathcal{T}_P \rightarrow \mathcal{T}_t$ can then be written as an invertible linear transformation which satisfies
\begin{equation}
{\mbf \lambda}_t = \phi_t {\mbf \lambda}_0,  \label{geom5}
\end{equation}
where ${\mbf \lambda}_t \in \mathcal{T}_t$ and ${\mbf \lambda}_0 \in \mathcal{T}_P$. Taking a derivative of \eqref{geom5} with respect to $t$, and using $\omega = -\dot{\phi}_t {\phi}_t^{-1}$ (superposed dot represent the derivative with respect to $t$), yields
\begin{equation}
\dot{\mbf \lambda}_t + \omega {\mbf \lambda}_t = {\bf 0}.  \label{geom6}
\end{equation}
If we assume $\omega$ to be a function of only $x^i(t)$ and $\dot{x}^i(t)$, then the theory of ordinary differential equations guarantee a unique solution to \eqref{geom6} for a given ${\mbf \lambda}_0$ at $t=0$. Invariance with respect to the choice of parameter requires $\omega$ to be homogeneous of degree one in $\dot{x}^i$. In particular, if $\omega$ is linear in $\dot{x}^i$ then it is called an {\it affine connection}; we will be concerned with only affine connections in this article. The components of $\omega$, with respect to the natural basis associated with $U$, are consequently given as
\begin{equation}
\omega^{i}_{j} = L^{i}_{jk} \dot{x}^k, \label{geom7}
\end{equation}
where $L^{i}_{jk}$ are twenty-seven real valued functions constituting the {\it connection coefficients}. Although $L^{i}_{jk}$ themselves do not form components of any tensor, the difference in two sets of connection coefficients does qualify for the components of a tensor. For example if $\hat{L}^{i}_{jk}$ is another set of connection coefficients, defined on the same coordinate neighborhood as $L^{i}_{jk}$, then the functions ${L}^{i}_{jk} - \hat{L}^{i}_{jk}$ are components of a tensor field.\footnote{cf. pp. 208-209 in Willmore, {\it op. cit.}} Given $L^{i}_{jk}$ we can introduce components of a symmetric connection $\Gamma^{i}_{jk}$  and a skew connection $T^{i}_{jk}$ such that
\begin{equation}
\Gamma^{i}_{jk} = \frac{1}{2}(L^{i}_{jk} + L^{i}_{kj}) ~~\text{and}~~  T^{i}_{jk} = \frac{1}{2}(L^{i}_{jk} - L^{i}_{kj}). \label{geom8}
\end{equation}
The latter set of functions form components of a tensor, called the {\it torsion tensor}, with respect to the natural basis over the coordinate neighborhood.

Let $\lambda^i$ denote the components of ${\mbf \lambda}_t$ with respect to the natural basis on $U$; they satisfy the following differential equations obtained from \eqref{geom6} and \eqref{geom7}:
\begin{equation}
\dot{\lambda}^i + L^{i}_{jk} \lambda^j \dot{x}^k  = 0.  \label{geom9}
\end{equation}
For a given set of connection coefficients the above equation has a unique solution with a prescribed initial value at $t=0$. The vector ${\mbf \lambda}_t$ is then said to be generated by {\it parallel transport} (or parallel displacement) of the vector ${\mbf \lambda}_0$ at $P$ along the curve $\gamma$. Equation \eqref{geom9} leads us to define the {\it intrinsic derivative} of a tangent vector along $\gamma$ as a vector whose components, denoted by $D\lambda^i/dt$, are given by
\begin{equation}
\frac{D\lambda^i}{dt} = \dot{\lambda}^i + L^{i}_{jk} \lambda^j \dot{x}^k.  \label{geom10}
\end{equation}
Hence the intrinsic derivative of ${\mbf \lambda}_t$, transported parallelly along $\gamma$, is identically zero. The intrinsic derivative of a scalar is identified with its normal derivative; i.e. for a scalar function $\phi$, defined on the curve $\gamma$,
\begin{equation}
\frac{D\phi}{dt} = \dot{\phi}.  \label{geom10.1}
\end{equation}
The intrinsic derivative of a tensor defined on $\gamma$, with components $a_{ij}$, is a tensor whose components, denoted by $Da_{ij}/dt$, can be determined by differentiating the scalar field $a_{ij}\lambda^i\lambda^j$ with respect to $t$ such that ${\mbf \lambda}_t$ is transported parallelly along $\gamma$. We obtain
\begin{equation}
\frac{Da_{ij}}{dt} = \dot{a}_{ij} - (L^{m}_{ik} a_{mj} + L^{m}_{jk} a_{im}) \dot{x}^k.  \label{geom10.2}
\end{equation}

If $\lambda^i$ are differentiable functions of the coordinates over $U$ then we can use the chain-rule to rewrite the right hand side of \eqref{geom10} as
\begin{equation}
\left( \lambda^i_{,k} + L^{i}_{jk} \lambda^j \right)\dot{x}^k,  \label{geom11}
\end{equation}
where $\lambda^i_{,k}$ stands for $\partial \lambda^i/\partial x_k$. The bracketed functions in \eqref{geom11} form components of a tensor, called the {\it covariant derivative} of ${\mbf \lambda}_t$. The components, denoted by $\lambda^i_{;k}$,  can be obtained from the following relations:
\begin{equation}
\lambda^i_{;k} = \lambda^i_{,k} + L^{i}_{jk} \lambda^j.  \label{geom12}
\end{equation}
Similarly, the covariant derivative of a tensor field, whose components $a_{ij}$ are (at least twice) differentiable functions of the coordinates over $U$, is a tensor, with components $a_{ij;k}$ given by
\begin{equation}
a_{ij;k} = a_{ij,k} - L^{m}_{ik} a_{mj} - L^{m}_{jk} a_{im}.  \label{geom13}
\end{equation}
If we assume the functions $L^{m}_{ik}$ to be differentiable then it is a matter of straightforward calculation to show that
\begin{equation}
a_{ij;kl} - a_{ij;lk} =  a_{hj} R^{h}_{ikl}  + a_{ih} R^{h}_{jkl} - 2 a_{ij;h} T^{h}_{kl},  \label{geom14}
\end{equation}
where the functions
\begin{equation}
R^{i}_{jkl} = L^{i}_{jl,k} - L^{i}_{jk,l} + L^{h}_{jl} L^{i}_{hk} - L^{h}_{jk} L^{i}_{hl} \label{geom15}
\end{equation}
are components of the {\it curvature tensor} (or the {\it Riemann-Christoffel tensor}) associated with the connection. Observe that the components are skew with respect to $k$ and $l$ index; hence there are only twenty-seven independent components of the curvature tensor. It is also clear that the covariant derivative of a tensor field is not commutative unless both the curvature and the torsion tensor are zero. In the next two subsections we prove that these conditions are in fact necessary and sufficient for the derivatives (of any order) to commute.

\subsection{Integrability of the affine connection}
 \begin{figure}[t!]
\centering
\subfigure[]{\includegraphics[scale=0.40]{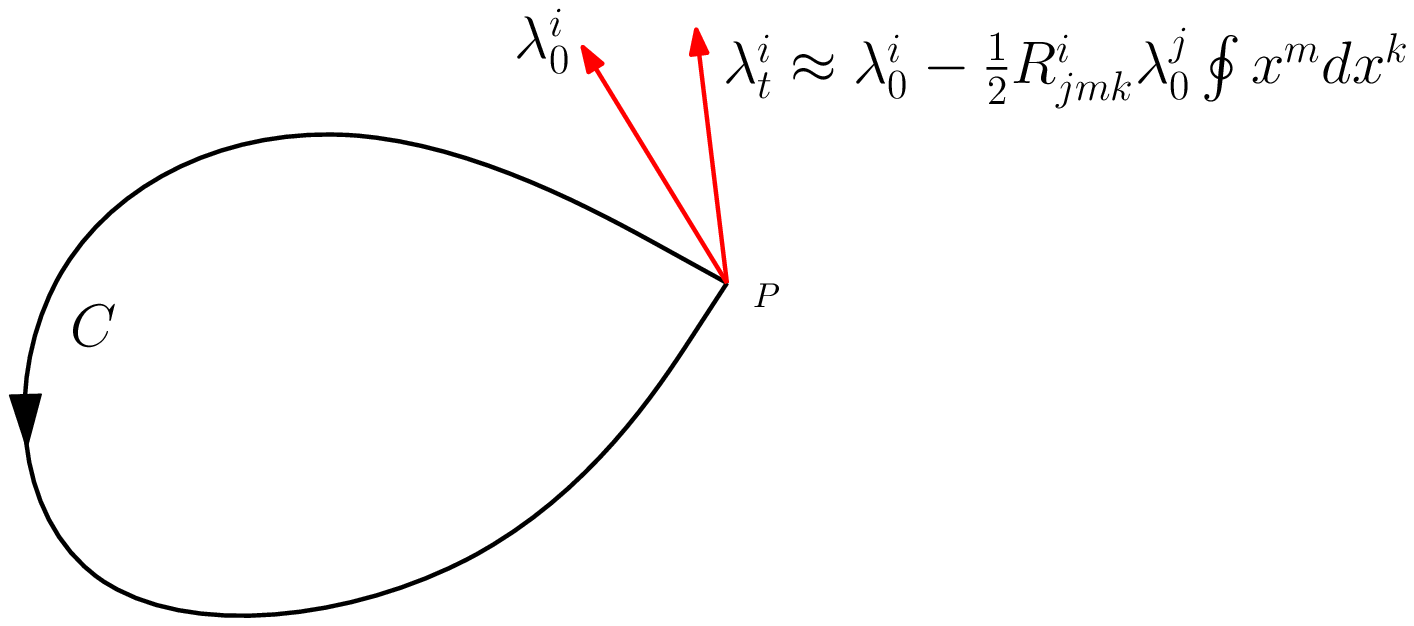}}
\hspace{10mm}
\subfigure[]{\includegraphics[scale=0.40]{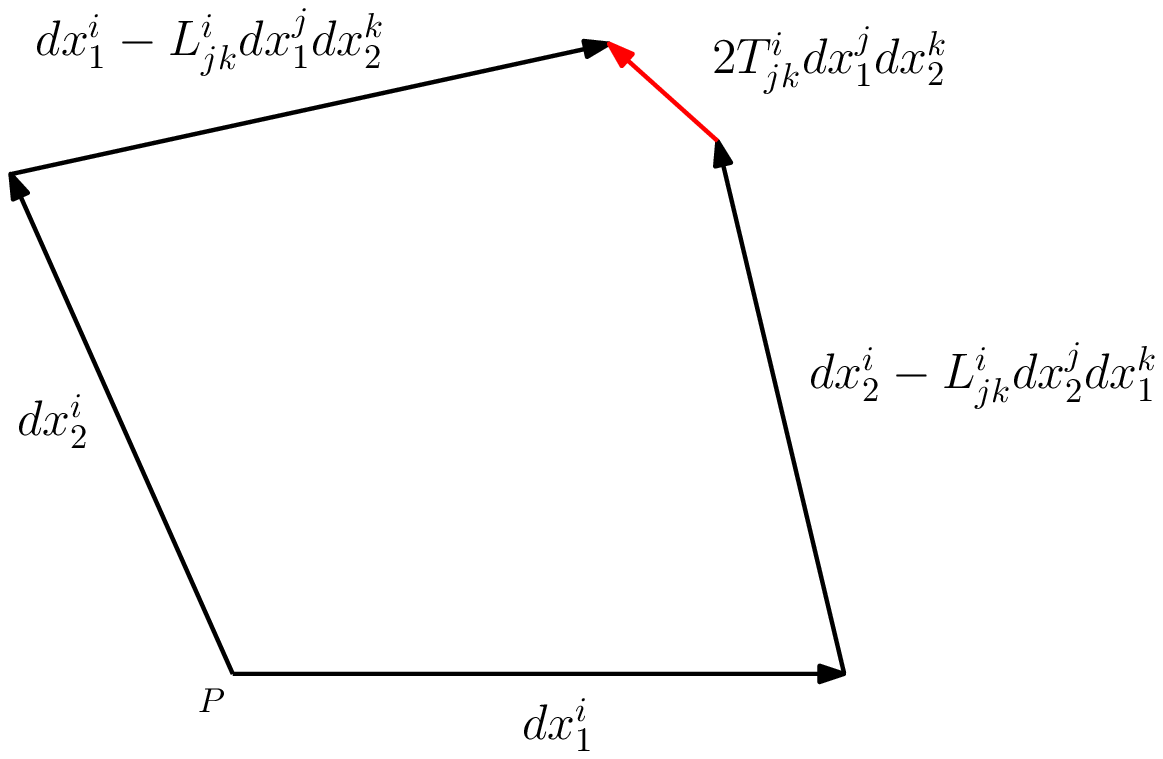}}
\caption{(a) Parallel displacement around a infinitesimally closed curve $C$. (b) Failure of closure of an infinitesimal parallelogram.}
\label{curvtor}
\end{figure}
Consider the parallel displacement of a tangent vector, initially at $P$ where its components are $\lambda^i_0$, along a curve $\gamma$ (as introduced above). The components of the vector at parameter value $t$, denoted by ${\lambda}^i_t$, are calculated by integrating \eqref{geom9} to get
\begin{equation}
{\lambda}^i_t = \lambda^i_0 - \int_0^t L^{i}_{jk}(x^p(\tau)) \lambda^j (\tau) \dot{x}^k (\tau) d\tau.  \label{integ0.1}
\end{equation}
The transported vector as obtained above will in general depend on the curve along which it has been displaced. In particular, a tangent vector when displaced parallelly around a closed curve will return to yield a different vector, see Figure \ref{curvtor}(a). The transport will be independent of the curve, and the vector upon parallel displacement around a closed curve will return to itself, if and only if the integrand in \eqref{integ0.1} can be written as a total derivative. The associated connection is then called {\it integrable}. A necessary and sufficient condition for the integrability of the connection is vanishing of the curvature tensor.\footnote{The following proof has be taken from pp. 135-138 in S. Weinberg, {\it Gravitation and Cosmology: Principles and Applications of the General Theory of Relativity}, John Wiley \& Sons, Inc., 1972; cf. p. 138 in Schouten, {\it op. cit.}} Indeed, consider an infinitesimally closed curve starting and ending at $P$, i.e. $x^p(t) = x^m(0)$. Both $L^{i}_{jk}(x^p(\tau))$ and $\lambda^j (\tau)$ can then be expanded around $x^p(0)$, within first order in $\left(x^m(\tau) - x^m(0)\right)$, as
\begin{eqnarray}
&& L^{i}_{jk}(x^p(\tau)) = L^{i}_{jk} (x^p(0)) +  \left(x^m(\tau) - x^m(0)\right) L^{i}_{jk,m}(x^p(0)) + \cdots ~ \text{and} \label{integ0.2}
\\
&& \lambda^i (\tau) = \lambda^i (0) - \left(x^k(\tau) - x^k(0)\right) L^{i}_{jk}(x^p(0)) \lambda^j (0) + \cdots, \label{integ0.3}
\end{eqnarray}
where \eqref{geom9} has been used in the latter. Substituting these into \eqref{integ0.1}, and retaining only the lowest order terms, we obtain
\begin{equation}
{\lambda}^i_t \approx \lambda^i_0 - \left( L^{i}_{jk,m} - L^{h}_{jm} L^{i}_{hk} \right) \lambda^j_0 \oint  {x}^m  d{x}^k.  \label{integ0.4}
\end{equation}
Owing to the fact that $\oint {x}^m  d{x}^k = - \oint {x}^k  d{x}^m$, and noting \eqref{geom15}, this can be written as
\begin{equation}
{\lambda}^i_t \approx \lambda^i_0 - \frac{1}{2} R^i_{jmk} \lambda^j_0 \oint  {x}^m  d{x}^k.  \label{integ0.5}
\end{equation}
Hence an arbitrary vector will not change upon being displaced parallelly around an infinitesimally closed curve if and only if $R^i_{jmk} = 0$. The same result will hold if we instead take an arbitrary closed curve within $U$. This is because the area enclosed by a closed curve can be broken down into several infinitesimal areas, bounded by infinitesimally closed curves, with the integral around the original closed curve given as a summation of integrals around all the infinitesimal closed curves.

Furthermore, for every integrable connection there exists an invertible matrix field with components $A^j_a$, whose inverse is given by $(A^{-1})^a_j$, such that\footnote{$(A^{-1})^a_jA^i_a = \delta^i_j$, where $\delta^i_j = 1$ when $i=j$ and $0$ otherwise.}
\begin{equation}
L^{i}_{jk} = -(A^{-1})^a_j A^i_{a,k} = A^i_a (A^{-1})^a_{j,k}. \label{integ1}
\end{equation}
To this end we note that path independence of parallel transport allows us to  unambiguously transport a vector parallelly to every other point in $U$; as a result we can construct a {\it parallel vector field}, also denoted by $\lambda^i$, such that $D\lambda^i/dt = 0$ for arbitrary curve.  It follows that the parallel vector field has a vanishing covariant derivative, i.e.
\begin{equation}
\lambda^i_{,k} + L^{i}_{jk} \lambda^j = 0. \label{integ2}
\end{equation}
Consider three linearly independent tangent vectors at $P$ with components $\lambda_1^i$, $\lambda_2^i$, and $\lambda_3^i$, respectively.\footnote{The three vectors are linearly independent when $c_1\lambda_1^i + c_2 \lambda_2^i + c_3\lambda_3^i = 0$ if and only if $c_1=c_2=c_3=0$, where $c_1, c_2$, and $c_3$ are scalars.} The parallel vector fields corresponding to each of these vectors satisfy
\begin{equation}
\lambda^i_{a,k} + L^{i}_{jk} \lambda_a^j = 0. \label{integ3}
\end{equation}
We can construct the connection functions from these relations by first identifying $\lambda^i_a$ with the components $A^i_a$ of a 3-by-3 matrix. The matrix is invertible due to linear independence of $\lambda^i_a$. Equations \eqref{integ1} now follow immediately from \eqref{integ3}; twenty-seven components of an integrable connection can therefore be derived only from nine functions of an invertible matrix field. In line with our expectations, the curvature tensor associated with the connection \eqref{integ1} is identically zero, as can be seen by substituting \eqref{integ1} into \eqref{geom15}.

One final remark regarding the integrability connections. Introduce $\mu^a = (A^{-1})^a_j \lambda^j$ and note that substituting \eqref{integ1}$_1$ into \eqref{integ2} reduces it to
\begin{equation}
\mu^a_{,k} = 0, \label{integ4}
\end{equation}
which imply that all the components $\mu^a$ are constant. This is essentially a feature of the components of a parallel vector field with respect to a Cartesian coordinate space. To elaborate, let $y^a$ be another system of coordinates used to describe points in $U$ such that the components of a tangent vector at a point, with respect to the two coordinate systems, are related as
\begin{equation}
dy^a = (A^{-1})^a_j dx^j. \label{integ5}
\end{equation}
It is immediately seen that the covariant derivative with respect to the $y^a$ coordinates are identical to the usual derivative; put differently, the components of the connection with respect to these new coordinates are identically zero. The coordinates $y^a$ are hence Cartesian in nature. However there is no bijective map between the two coordinate systems unless $A^j_a$ is integrable, i.e. unless $A^j_a = \partial x^j /\partial y^a$.

\label{affine}

\subsection{Torsion tensor} The torsion tensor related to the connection is related to the failure of drawing infinitesimal parallelograms at the point, see \eqref{geom8}$_2$ and Figure \ref{curvtor}(b).\footnote{cf. p. 127 in Schouten, {\it op. cit.}}  Two tangent vectors $dx_1^i$ and $dx_2^i$ at $x^k$ (coordinates of $P$) when displaced parallelly along each other become $dx_1^i - L^i_{jk} dx_1^j dx_2^k$ and $dx_2^i - L^i_{jk} dx_2^j dx_1^k$ at $x^k + dx_2^k$ and $x^k + dx_1^k$, respectively while neglecting terms of higher order, cf. \eqref{integ0.3}. The resulting figure fails to close, and thus fails to create a parallelogram, with the closing vector given by, see Figure \ref{curvtor}(b),
\begin{equation}
2 T^{i}_{jk} dx_1^j dx_2^k. \label{tor1}
\end{equation}
In a space with symmetric connection, the torsion will vanish identically and hence infinitesimal parallelograms will exist at every point.

Since $T^{i}_{jk}$ is skew with respect to the $j$ and $k$ index, there always exist a tensor density ${\mbf \alpha}$ whose components $\alpha^{mi}$ are defined by\footnote{The permutation symbol $e^{ijk}$ is such that $e^{ijk} = 0$ if any two indices are same, $e^{ijk} = 1$ if $ijk$ is an even permutation of $123$, and $e^{ijk} = -1$ if $ijk$ is an odd permutation of $123$. The components $e_{ijk}$ are defined similarly. Below are some useful identities involving these symbols:
\begin{eqnarray}
&& e^{ijk} (A^{-1})^a_i (A^{-1})^b_j (A^{-1})^c_k = \det (A^{-1}) e^{abc}, \label{torfoot1}
\\
&& e^{ijk} e_{ijk} = 6, ~e^{ijk} e_{ijr} = 2 \delta^{k}_{r},~e^{ijk} e_{iqr} = \left| \begin{array}{cc}
                                                                                        \delta^{j}_{q} & \delta^{j}_{r} \\
                                                                                        \delta^{k}_{q} & \delta^{k}_{r}
                                                                                      \end{array}
\right|, ~e^{ijk} e_{pqr} = \left| \begin{array}{ccc}
                                                                                        \delta^{i}_{p} & \delta^{i}_{q} & \delta^{i}_{r} \\
                                                                                        \delta^{j}_{p} & \delta^{j}_{q} & \delta^{j}_{r} \\
                                                                                        \delta^{k}_{p} & \delta^{k}_{q} & \delta^{k}_{r}
                                                                                      \end{array}
\right|, \label{torfoot2}
\end{eqnarray}
where the notation $|\cdot|$ represents the determinant of the contained matrix. For proofs etc. see pp. 100-108 in I. S. Sokolnikoff, {\it Tensor Analysis}, John Wiley \& Sons, Inc., 1951.}
\begin{equation}
\alpha^{mi} = e^{mjk} T^{i}_{jk}.  \label{tor3}
\end{equation}
On the other hand it follows from \eqref{integ1} that, for an integrable connection, the associated torsion tensor can be written as
\begin{equation}
2T^{i}_{jk} =  (A^{-1})^a_k A^i_{a,j} -(A^{-1})^a_j A^i_{a,k} = (A^{-1})^a_k (A^{-1})^b_j \frac{\partial A^i_a}{\partial y^b} -(A^{-1})^a_j (A^{-1})^b_k \frac{\partial A^i_a}{\partial y^b}, \label{tor2}
\end{equation}
where the second equality has been obtained using \eqref{integ5}. Substituting from \eqref{tor2}$_2$ and using the identity \eqref{torfoot1}, we get
\begin{equation}
\alpha^{mi} = \det (A^{-1}) e^{cba}  A^m_c \frac{\partial A^i_a}{\partial y^b} = \det (A^{-1})  A^m_c (\Curl A)^{ci},  \label{tor4}
\end{equation}
where $(\Curl A)^{ci} \coloneqq e^{cba} \partial A^i_a / \partial y^b$ are the components of the curl of a tensor. Going back to \eqref{integ5} we can now state that $U$, with an integrable connection, is diffeomorphic to a rectilinear (or Cartesian) space if and only if the torsion field in $U$ vanishes identically.

\label{torsion}

\subsection{Riemannian geometry} Consider a differentiable manifold $\mathcal{X}$, as introduced above, with tangent spaces equipped with an inner product given by a bilinear form. Thus, for an arbitrary point $P$ on the manifold with a coordinate neighborhood $U$, we have an inner product given by
\begin{equation}
g_{ij} \lambda^i \mu^j, \label{riem1}
\end{equation}
where $\lambda^i$ and $\mu^j$ are components of two tangent vectors ${\mbf \lambda}$ and ${\mbf \mu}$ in $\mathcal{T}_P$ with respect to the natural basis associated with $U$; $g_{ij}$ are components of a tensor field ${\bf g}$ with respect to the natural basis on U. A differentiable manifold with the inner product \eqref{riem1}, where ${\bf g}$ is assumed to be non-singular, symmetric, and (at least twice) differentiable, is called a {\it Riemannian Manifold}. We assume, in addition, that ${\bf g}$ is positive-definite. The magnitude of ${\mbf \lambda}$, denoted by $|{\mbf \lambda}|$, can therefore be defined by the following quadratic form:
\begin{equation}
|{\mbf \lambda}|^2 = g_{ij} \lambda^i \lambda^j. \label{riem2}
\end{equation}
The tensor ${\bf g}$ is called the {\it metric} tensor. The angle $\theta$ between two tangent vectors ${\mbf \lambda}$ and ${\mbf \mu}$ is given by
\begin{equation}
\cos \theta = \frac{g_{ij} \lambda^i \mu^j}{|{\mbf \lambda}||{\mbf \mu}|}. \label{riem3}
\end{equation}
The manifold is Euclidean if all the points on the manifold can be covered with a single coordinate neighborhood and there exists a coordinate system, associated with the coordinate neighborhood, with respect to which $g_{ij} = 1$ for $i = j$ and $0$ otherwise for all points on the manifold. The Riemmanian manifold is essentially Euclidean in each of its infinitesimal parts. It should also be emphasized that a three-dimensional Riemannian manifold cannot always be embedded in a four-dimensional Euclidean space; this was assumed earlier for the two-dimensional curved surface. An example of a two-dimensional manifold which cannot be embedded in a three-dimensional Euclidean space is given by the Klein bottle.

According to the fundamental theorem of Riemannian geometry, for a given Riemannian metric there exists a unique symmetric affine connection (called the Riemannian connection or the Levi-Civita connection) which transports vectors parallelly while preserving the inner product; as a result both the length and the angle between vectors are preserved during parallel transport.\footnote{cf. p. 228 in Willmore, {\it op. cit.}} Accordingly, given a metric field with components $g_{ij}$ there exists a symmetric connection $\Gamma^i_{jk}$ (i.e. $\Gamma^i_{jk} = \Gamma^i_{kj}$) which preserves the inner product during parallel transport along a curve, say $\gamma$ (see the earlier discussion regarding $\gamma$), i.e.
\begin{equation}
\frac{D}{dt} g_{ij} \lambda^i \mu^j = 0 \label{riem4}
\end{equation}
along $\gamma$ such that $D\lambda^i/{dt} = D\mu^i/{dt} = 0$. Given the arbitrariness of the curve, as well as of the vectors, \eqref{riem4} requires vanishing of the covariant derivative of the metric tensor, i.e.
\begin{equation}
g_{ij|k} = 0. \label{riem5}
\end{equation}
The symbol $|$ in the subscript is used to denote the covariant derivative associated with the Riemannian connection, so as to distinguish it from the covariant derivative with respect to a general connection. The above condition, when used in \eqref{geom13} with $g_{ij}$ in place of $a_{ij}$ and $\Gamma^i_{jk}$ in place of $L^{i}_{jk}$, after some straightforward manipulation yields
\begin{equation}
\Gamma^k_{ij} = g^{hk} [h~ij], \label{riem6}
\end{equation}
where $g^{hk}$ satisfy $g^{hk}g_{kj} = \delta^h_j$ and $[h~ij] \coloneqq 1/2 \left( g_{ih,j} + g_{jh,i} - g_{ij,h} \right)$ is called the Christoffel symbol of first kind, cf. \eqref{geom3.1}. The symmetric connection $\Gamma^k_{ij}$ is often denoted by $\left\{\begin{smallmatrix} k \\ ji  \end{smallmatrix} \right\}$ and is called the Christoffel symbol of the second kind.

The components of the Riemann-Christoffel curvature tensor associated with the Riemannian connection, denoted by $K^{i}_{jkl}$, can be calculated by substituting \eqref{riem6} into \eqref{geom15}. We obtain, for $K_{hjkl} = g_{hi} K^{i}_{jkl}$,
\begin{equation}
K_{mjkl} = \frac{1}{2} \left( g_{lm,jk} - g_{km,jl} + g_{jk,lm} - g_{jl,km} \right) + [i~ml] \left\{\begin{smallmatrix} i \\ jk  \end{smallmatrix} \right\} - [i~mk] \left\{\begin{smallmatrix} i \\ jl  \end{smallmatrix} \right\}. \label{riem7}
\end{equation}
The following symmetries are evident from the above formula: i) $K_{mjkl} = K_{klmj}$, ii) $K_{mjkl} = -K_{jmkl}$, $K_{mjkl} = -K_{mjlk}$. They reduce the number of independent components of the curvature tensor from $81$ to $6$ (in two-dimensions there is only one independent component, the Gaussian curvature). In fact, in the present three-dimensional settings, the curvature tensor can be completely defined in terms of a second-order symmetric tensor (known as Einstein tensor) whose components are given by\footnote{The Einstein tensor is generally introduced in terms of the symmetric Ricci tensor, $R_{jk} \coloneqq K^{i}_{jki}$, as $E^{pq} = R^{pq} - \frac{1}{2} g^{pq} R$, where $R^{pq} = g^{pj} g^{qk} R_{jk}$ and $R = g^{jk} R_{jk}$. That this definition is identical to \eqref{riem8} in three-dimensions can be shown by using the identity
\begin{equation}
e^{ijk} e^{lmn} = g g^{lp} g^{mq} g^{nr} \left| \begin{array}{ccc}
                                                                                        \delta^{i}_{p} & \delta^{i}_{q} & \delta^{i}_{r} \\
                                                                                        \delta^{j}_{p} & \delta^{j}_{q} & \delta^{j}_{r} \\
                                                                                        \delta^{k}_{p} & \delta^{k}_{q} & \delta^{k}_{r}
                                                                                      \end{array}
\right|, \label{riemfoot1}
\end{equation}
which can be proved using $e_{ijk} g^{ai} g^{bj} g^{ck} = \det (g^{pq}) e^{abc}$, \eqref{torfoot1}, and \eqref{torfoot2}$_4$.}
\begin{equation}
E^{pq} = \frac{1}{4g} e^{pmj} e^{qkl} K_{mjkl}, \label{riem8}
\end{equation}
where $g = \det (g_{ij})$. The above relation can be inverted, using \eqref{torfoot2}$_3$, to get
\begin{equation}
K_{mjkl} = g e_{pmj} e_{qkl} E^{pq}. \label{riem9}
\end{equation}
The curvature tensor additionally satisfies the Bianchi's identity
\begin{equation}
K^i_{jkl|h} + K^i_{jhk|l} + K^i_{jlh|k} = 0, \label{riem9.1}
\end{equation}
which can be verified by direct substitution.\footnote{cf. pp. 144-152 in Schouten, {\it op. cit.}} Contracting the indices $i$ and $l$, followed by some straightforward manipulation, yields the Einstein tensor to be divergence free, i.e.
\begin{equation}
E^{pq}_{~~|q} = 0. \label{riem10}
\end{equation}
This identity plays a central role in the relativity theory.

If the curvature tensor given in \eqref{riem7}, or equivalently the Einstein tensor, vanishes identically at all points in the manifold then the connection $\left\{\begin{smallmatrix} i \\ jk  \end{smallmatrix} \right\}$ is necessarily of the form \eqref{integ1}. Moreover since the torsion tensor associated with this connection is zero the tensor $A^i_a$ in \eqref{integ1} is integrable; then there exists Cartesian coordinates $y^a$, related bijectively to $x^i$, such that
\begin{equation}
\left\{\begin{smallmatrix} i \\ jk  \end{smallmatrix} \right\} = \frac{\partial x^i}{\partial y^a} \frac{\partial^2 y^a}{\partial x^j \partial x^k}~\text{and}~g_{ij} = \frac{\partial y^a}{\partial x^i} \frac{\partial y^b}{\partial x^j} \delta_{ab}, \label{riem11}
\end{equation}
cf. \eqref{tor4}, \eqref{integ5}, and the discussion which follows them. The Riemannian manifold in such a case reduces to a Euclidean space.

\subsection{Further generalizations} Knowing the metric function is sufficient to determine the local geometric structure in a Riemannian space. This is not so if instead of the Riemannian connection \eqref{riem6} we consider a connection, denoted by $L^{i}_{jk}$, which is not necessarily symmetric and with respect to which the covariant derivative of $g_{ij}$ does not vanish. There are three important tensors associated with this connection: (i) the torsion tensor whose components $T^{i}_{jk}$ are derived from the skew part of the connection, see \eqref{geom8}$_2$, (ii) the nonmetricity tensor whose components $Q_{kij}$ are defined as\footnote{The concept of nonmetricity was introduced by Hermann Weyl, cf. pp. 121-125 in Weyl, {\it op. cit.} in an attempt to unify gravity with electromagnetism. He assumed the nonmetricity to be of the form $Q_{kij} = Q_i g_{jk}$, where $Q_i$ are components of a vector. Such a form of the nonmetricity preserves the ratio of the magnitude of two vectors during parallel displacement. The present definition of nonmetricity, i.e. \eqref{friem1}$_1$, is taken from pp. 131-132 in Schouten, {\it op. cit.} \label{weyl}}
\begin{equation}
Q_{kij} \coloneqq - g_{ij;k} = -g_{ij,k} + L^{m}_{ik} g_{mj} + L^{m}_{jk} g_{im}, \label{friem1}
\end{equation}
where the second equality follows from \eqref{geom13}, and (iii) the curvature tensor $R^i_{jkl}$ associated with the connection, see \eqref{geom15}. Note that the nonmetricity tensor is symmetric with respect to the last two indices.

Construct two equations from \eqref{friem1} by permutating the indices $kij$ as $kji$ and $ijk$. Adding the former to \eqref{friem1}, and subtracting the later from it, we obtain\footnote{cf. p. 132 in Schouten, {\it op. cit.} and p. 125 in Weyl, {\it op. cit.} The latter provides an instance of a symmetric affine connection in a space with nonmetricity of the form given in the previous footnote. Hermann Weyl, in fact, was the first to introduce the idea of a general affine connection not necessarily derived from a metric. The notion of affine connection was further generalized by \`{E}lie Cartan who introduced the concept of torsion tensor, among various other contributions to differential geometry, cf. E. Cartan, {\it On Manifolds With an Affine Connection and the Theory of General Relativity} (translated by A. Magnon and A. Ashtekar), Bibliopolis, 1986.}
\begin{equation}
L^{n}_{jk}  = \left\{\begin{smallmatrix} n \\ jk  \end{smallmatrix} \right\} + T^{n}_{jk} -  g^{in} (T^{m}_{ik} g_{jm} + T^{m}_{ij} g_{km}) + \frac{1}{2} g^{in} (Q_{kij} + Q_{jki} - Q_{ijk}). \label{friem2}
\end{equation}
An affine connection can therefore be constructed using metric, torsion, and nonmetricity tensors. Clearly a connection which is required to be symmetric, i.e. with vanishing torsion, and which preserve inner products during parallel transport, i.e. has vanishing nonmetricity, is necessarily identical to the Riemannian connection. This is expected following the fundamental theorem of Riemannian geometry as stated in the previous subsection. On the other hand, requiring $L^{n}_{jk}$ to be symmetric yields a zero torsion but places no limitation of the nonmetricity. A differential manifold with a connection given by \eqref{friem2} give rise to what is called the Riemann-Cartan-Weyl geometry.

We introduce two tensors, with components defined by
\begin{eqnarray}
&& K^n_{jk} \coloneqq T^{n}_{jk} -  g^{in} (T^{m}_{ik} g_{jm} + T^{m}_{ij} g_{km})~\text{and} \label{friem3}
\\
&& M^n_{jk} \coloneqq \frac{1}{2} g^{in} (Q_{kij} + Q_{jki} - Q_{ijk}). \label{friem4}
\end{eqnarray}
The former is called the contortion tensor. If $W^n_{jk} \coloneqq K^n_{jk} + M^n_{jk}$ then we can write \eqref{friem2} as
\begin{equation}
L^n_{jk} = \left\{\begin{smallmatrix} n \\ jk  \end{smallmatrix} \right\} + W^n_{jk}. \label{friem5}
\end{equation}
Substituting this into \eqref{geom15} we get components of the curvature tensor as\footnote{cf. p. 141 in Schouten, {\it op. cit.}}
\begin{equation}
R^i_{jkl} = K^i_{jkl} + W^i_{jl|k} - W^i_{jk|l} + W^h_{jl}W^i_{hk} - W^h_{jk}W^i_{hl}, \label{friem6}
\end{equation}
where recall that the subscript $|$ denotes the covariant derivative associated with the Riemannian connection. This relation should be seen as the central result in this brief excursion on differential geometry. It is most useful, at least for purposes in elasticity theory, when seen as a nonlinear partial differential equation for the metric functions $g_{ij}$ for given curvature, torsion, and nonmetricity tensors.

We end our present discussion by mentioning several identities associated with $R^i_{jkl}$.\footnote{cf. pp. 144-152 in Schouten, {\it op. cit.}} (a) As noted earlier following \eqref{geom15}, $R^i_{jkl}$ is skew with respect to $k$ and $l$ index. Hence unlike the Riemann curvature tensor $K^i_{jkl}$, which has only six independent components, it has twenty-seven independent components. (b) For any tensor with components $D_{jkl}$ denote $6 D_{[jkl]} = (D_{jkl} + D_{klj} + D_{ljk} - D_{lkj} - D_{jlk} - D_{kjl})$ as its skew alternation with respect to all the indices. Using \eqref{geom15} repeatedly we can then obtain
\begin{equation}
R^i_{[jkl]} = 2 T^i_{[lk;j]} - 4 T^h_{[kl} T^i_{j]h}. \label{friem7}
\end{equation}
(c) Recall \eqref{geom14}, with $g_{ij}$ in place of $a_{ij}$, and use \eqref{friem1}$_1$ to write
\begin{equation}
R_{ijkl} + R_{jikl} = Q_{kij;l} - Q_{lij;k} + 2 Q_{hij} T^h_{kl}. \label{friem8}
\end{equation}
Hence for a vanishing nonmetricity $R_{ijkl}$ is skew with respect to the first two indices. (d) The Bianchi's identity
\begin{equation}
R^i_{j[kl;m]} = 2  R^i_{jh[l} T^h_{mk]}, \label{friem9}
\end{equation}
which can be verified by direct substitution. For a symmetric connection it reduces down to a form previously written for $K^i_{jkl}$ in \eqref{riem9.1}. The above identities are important relations between curvature tensor, torsion, and nonmetricity. In the following sections we will relate these tensors with various types of defect densities.

\section{Isolated defects in crystalline solids}
Crystalline solids constitute a class of ordered media whose microstructure appears as a highly regular pattern of lattice points. They are in general replete with local anomalies or {\it defects} which destroy the crystalline order. Defects can be present both naturally in a crystal or can appear as a result of external influence, e.g. during thermal activation, irradiation, or plastic deformation. A defective solid can have significantly different physical properties (mechanical, electrical, chemical, optical etc.) in comparison to an ideal solid; for instance, the plastic nature of metals is essentially governed by the mechanics of dislocations and grain boundaries, and the optical nature of liquid crystals by the mechanics of disclinations.

In a three-dimensional crystalline solid body, defects can be broadly classified based on their dimensionality. Therefore, we have zero-dimensional defects (or point defects) in the form of vacancies (missing atoms), interstitials (extra atoms of the same kind), and substitutionals (extra atoms of a different kind), see Figure \ref{pdefect1}; one-dimensional defects in the form of dislocations and disclinations; two-dimensional defects such as grain boundaries, phase boundaries, domain walls, stacking faults, and free surfaces; and three dimensional defects in the form of precipitates and inhomogeneities.\footnote{For a general treatment of point defects in solids see A. M. Stoneham, {\it Theory of Defects in Solids}, Oxford, 1975; for distortion and stress fields associated with point defects, as well as for precipitates and inhomogeneities, see the pertinent papers of J. D. Eshelby, e.g. in Markenscoff and Gupta, {\it op. cit.} For an extensive application of dislocation theory to diverse areas of physics, see F. R. N. Nabarro, {\it Theory of Crystal Dislocations}, Dover, 1987. For disclinations and other defects in liquid crystals and magnetic media see M. Kl\'{e}man, {\it Points, Lines and Walls}, John Wiley \& Sons, 1983. For an excellent perspective on the theory and application of defects from varied disciplines of mathematics (topology, geometry), physics (condensed matter, statistical mechanics), geology (glacier flow, earthquakes), and biology (membranes, cells) see the rich collection of expository articles in R. Balian et. al. (Eds.), {\it Les Houches, Session XXXV, 1980 -- Physique des d{\'e}fauts (Physics of Defects)}, North-Holland, 1981.}

In the present section our purpose is to illustrate zero and one-dimensional defects in a two-dimensional Bravais lattice and subsequently motivate their non-Euclidean character.\footnote{cf. A.-H. Anthony, {\it Sol. St. Phen.}, 87, pp. 15-46, 2002.} In particular, with simple geometric illustrations, we will associate the torsion tensor, the curvature tensor, and the nonmetricity tensor with the presence of dislocations, disclinations, and point defects, respectively. The defects also render the crystal internally stressed. In each of the illustrations, defects are introduced by disturbing the minimum energy configuration of the perfectly ordered crystal; this has been done by either adding or removing an extra atomic half-plane, a wedge, and a single particle (of same or different constitution) into the ordered lattice and then joining of the open cuts afterwards to create a dislocation, disclination, and a point defect, respectively. These surgeries immediately give rise to the internal stress fields, much like permanent damage. To relax the defective crystal from its internally stressed state, it has to be cut into pieces; for instance a single cut suffices to relax the dislocated crystal in Figure \ref{discubic1}(b). These pieces, in general, do not fit together to form a continuous crystal in a three-dimensional Euclidean space. They, however, would fit together in some non-Euclidean space. This can be seen clearly by considering a single dislocation inside a plate. If the plate is constrained to remain flat even after introducing the dislocation then it would develop internal stresses; however if left unconstrained the plate would bend to come to a natural or stress free state. Hence to relax stresses the plate has to necessary leave the two-dimensional Euclidean space and occupy a two-dimensional Riemannian space embedded in a three-dimensional space. We will use these insights to develop, in the following section, a general theory of continuous distribution of defects in solids. When distributed continuously, the defects do not exist in isolation but instead are smeared over to give rise to effective fields of defect densities.

We do not make any attempt to provide even an elementary introduction to defect theory in the following subsections, something which can be easily accessed from several excellent texts, but instead we restrict ourselves to demonstrate their non-Euclidean nature.

% In more formal terms, we are interested in characterizing the geometry of a two-dimensional manifold whose points are identified with the lattice points, and the tangent vectors at each lattice point with the arrows which connect it to the lattice points in the immediate neighborhood. A metric can be introduced with which changes in length of a tangent vector, as well as the angle between two of them, can be measured with respect to the perfect lattice. We will associate two affine connections with the manifold. Recall that affine connections are linear maps which take tangent vectors at a lattice point and displace them `parallelly' to another lattice point. One is the Riemmanian connection derived from the metric, which by construction has zero torsion and nonmetricity tensors.     A perfect Bravais lattice (i.e. without any defects) has a trivial affine connection associated with it. Otherwise, it is the nature of the defect (or the group of defects) which determines the form of the affine connection.  The two-dimensional manifold is what defines the {\it material space} associated with the body. The affine connection is called {\it material connection} so as to distinguish it from the Riemannian connection of the metric. Finally we note that all our geometric considerations are for points away from the core of the defect whose singular nature makes it otherwise difficult to define above notions in an unambiguous fashion.

\label{lattice}

\subsection{Dislocation}
\begin{figure}[t!]
\centering
\subfigure[A perfect cubic Bravais crystal]{\includegraphics[scale=0.25]{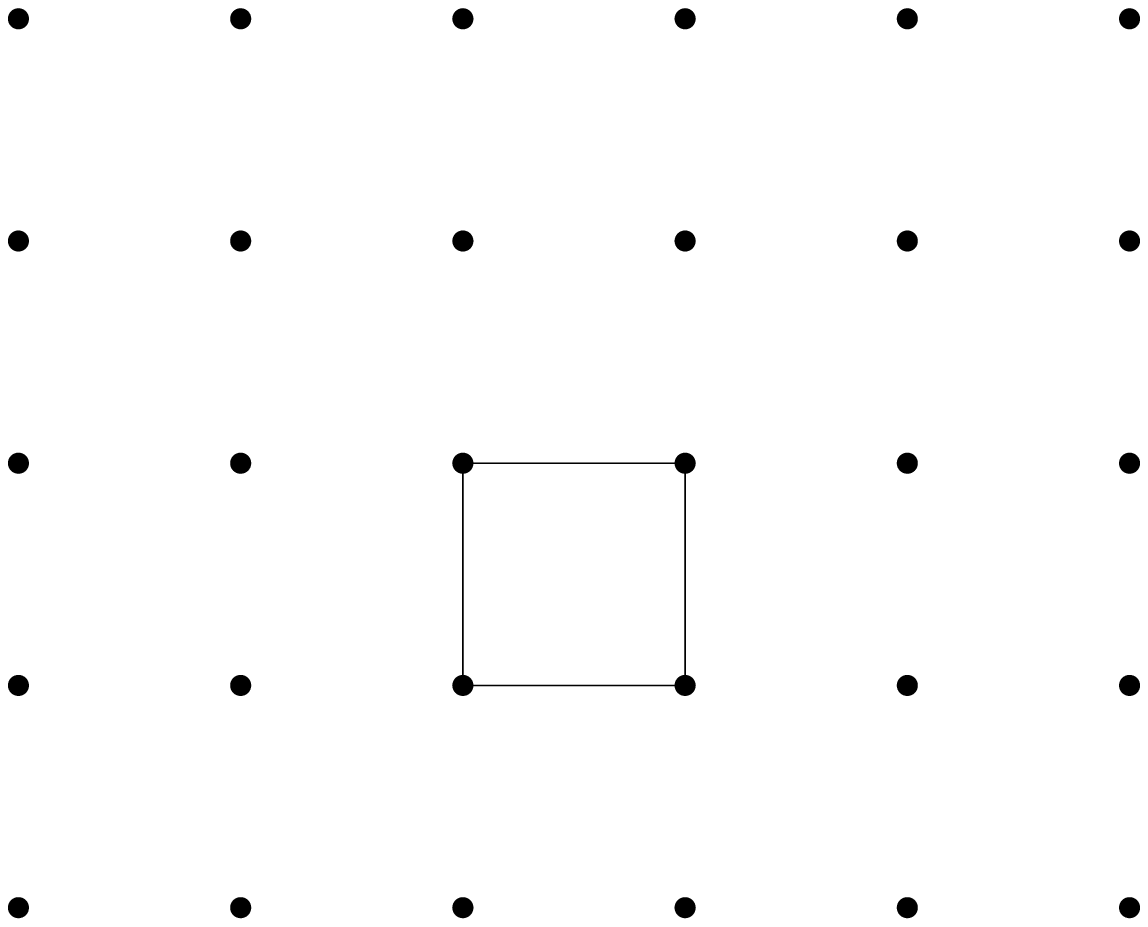}}
\hspace{6mm}
\subfigure[A positive dislocation]{\includegraphics[scale=0.25]{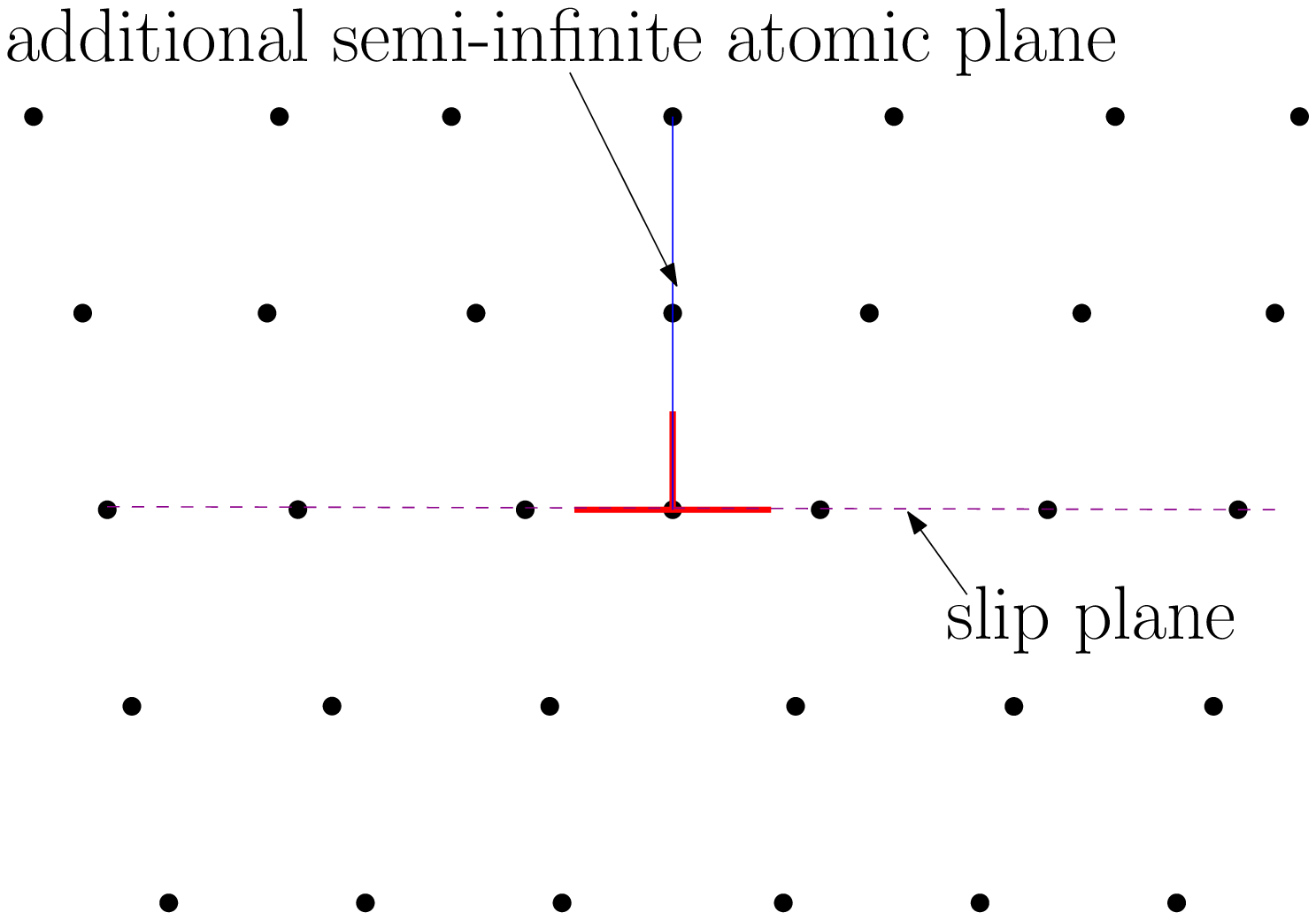}}
\hspace{6mm}
\subfigure[A perfect hexagonal Bravais crystal]{\includegraphics[scale=0.35]{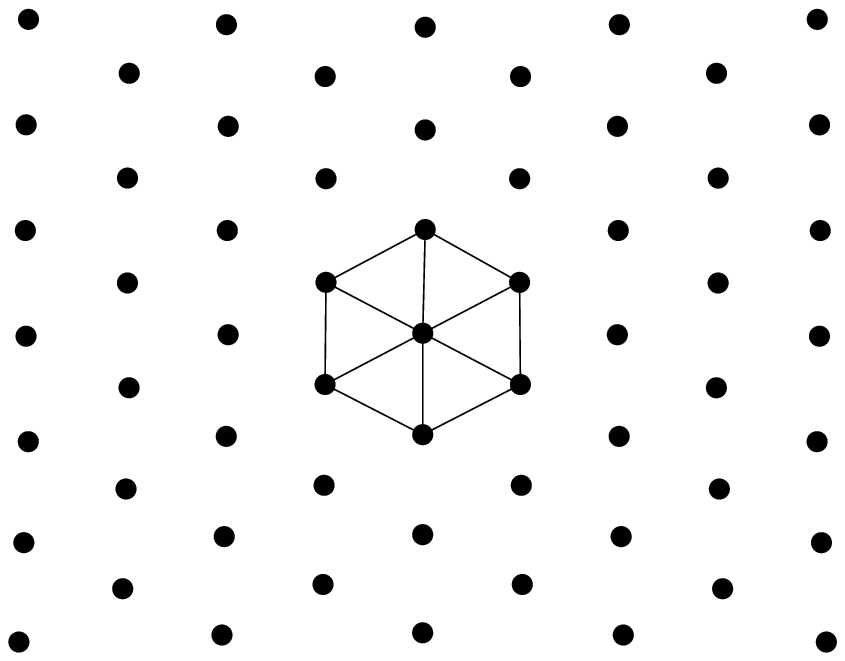}}
\hspace{6mm}
\subfigure[A negative dislocation]{\includegraphics[scale=0.35]{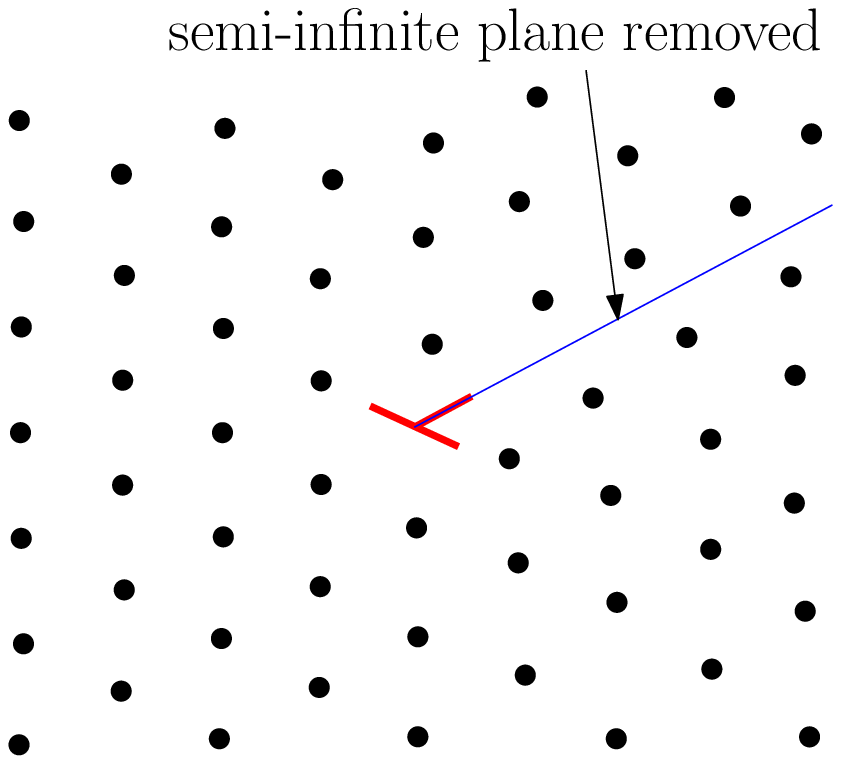}}
\caption{Formation of a positive and a negative dislocation in a cubic and a hexagonal Bravais crystal, respectively.}
\label{discubic1}
\end{figure}
Dislocations are the most important defects in a crystal; they govern not only the deformation and the strength of crystalline solids but also their growth behavior. A dislocation can be created by inserting (or removing) a planar array of atoms in a perfect lattice. For example in perfect Bravais crystals shown in Figures \ref{discubic1}(a) and \ref{discubic1}(c) a semi-infinite atomic plane has been inserted and removed, respectively. The edge of this plane, which is a straight line piercing in and out of the paper ad infinitum (seen as a point in these figures), is identified as a linear defect called the dislocation.

\begin{figure}[t!]
\centering
\includegraphics[scale=0.30]{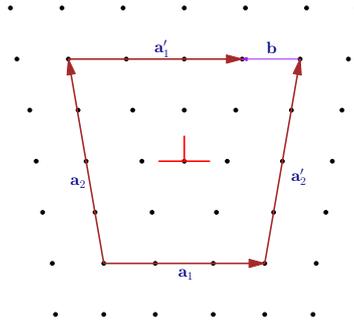}
\caption{Closure failure of a parallelogram in a dislocated crystal.}
\label{discubic2}
\end{figure}

A dislocation is essentially a translational defect; its introduction into the lattice, in effect, translates all the lattice points above the `slip' plane by one lattice spacing (this is clearly visualized when the crystal is cut into two halves along the slip plane). In doing so, it maintains both the orientation of the lattice as well as the size of lattice spacing. These facts are now illustrated geometrically. We start by drawing a parallelogram around the dislocation, see Figure \ref{discubic2}. Two tangent vectors ${\bf a}_1$ and ${\bf a}_2$, at a lattice point close to the dislocation line, are transported parallelly along ${\bf a}_2$ and ${\bf a}_1$ to yield tangent vectors ${\bf a}_1^\prime$ and ${\bf a}_2^\prime$, respectively. The parallelogram formed by these four vectors is open. The deficiency, denoted as ${\bf b}$ in Figure \ref{discubic2}, is called the Burgers vector of the dislocation. A parallelogram constructed around a regular lattice point, away from the dislocation, will always close. Drawing an analogy with our earlier discussion in Subsection \ref{torsion} we are thus led to associate dislocation with the torsion in a geometric space. On the other hand the fact that a dislocation preserves both the orientation and the spacing of the lattice has following geometrical consequences. Any tangent vector, when translated parallelly along a closed curve around the dislocation, remains invariant; thus the affine connection, which otherwise has a non-zero torsion, is integrable, cf. \ref{affine}. Secondly, the length of a tangent vector remains unchanged as it traverses the lattice forcing the covariant derivative of the metric, with respect to the affine connection, to vanish identically. To summarize, the affine connection related to dislocation has a non-vanishing torsion but zero curvature and nonmetricity.

\subsection{Disclination}
\begin{figure}[t!]
\centering
\subfigure[Removal of a $90^\circ$ wedge from a cubic Bravais crystal]{\includegraphics[scale=0.34]{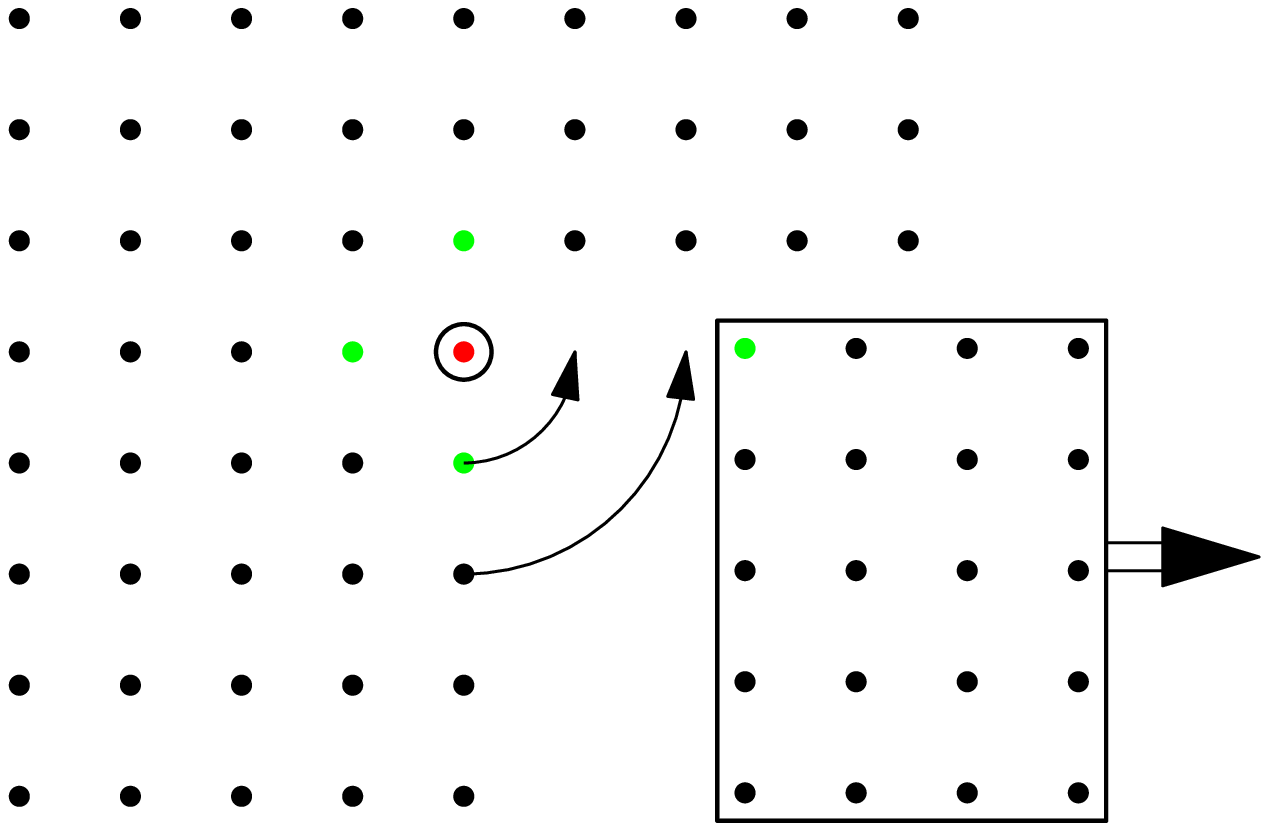}}
\hspace{1mm}
\subfigure[A -$90^\circ$ disclination]{\includegraphics[scale=0.18]{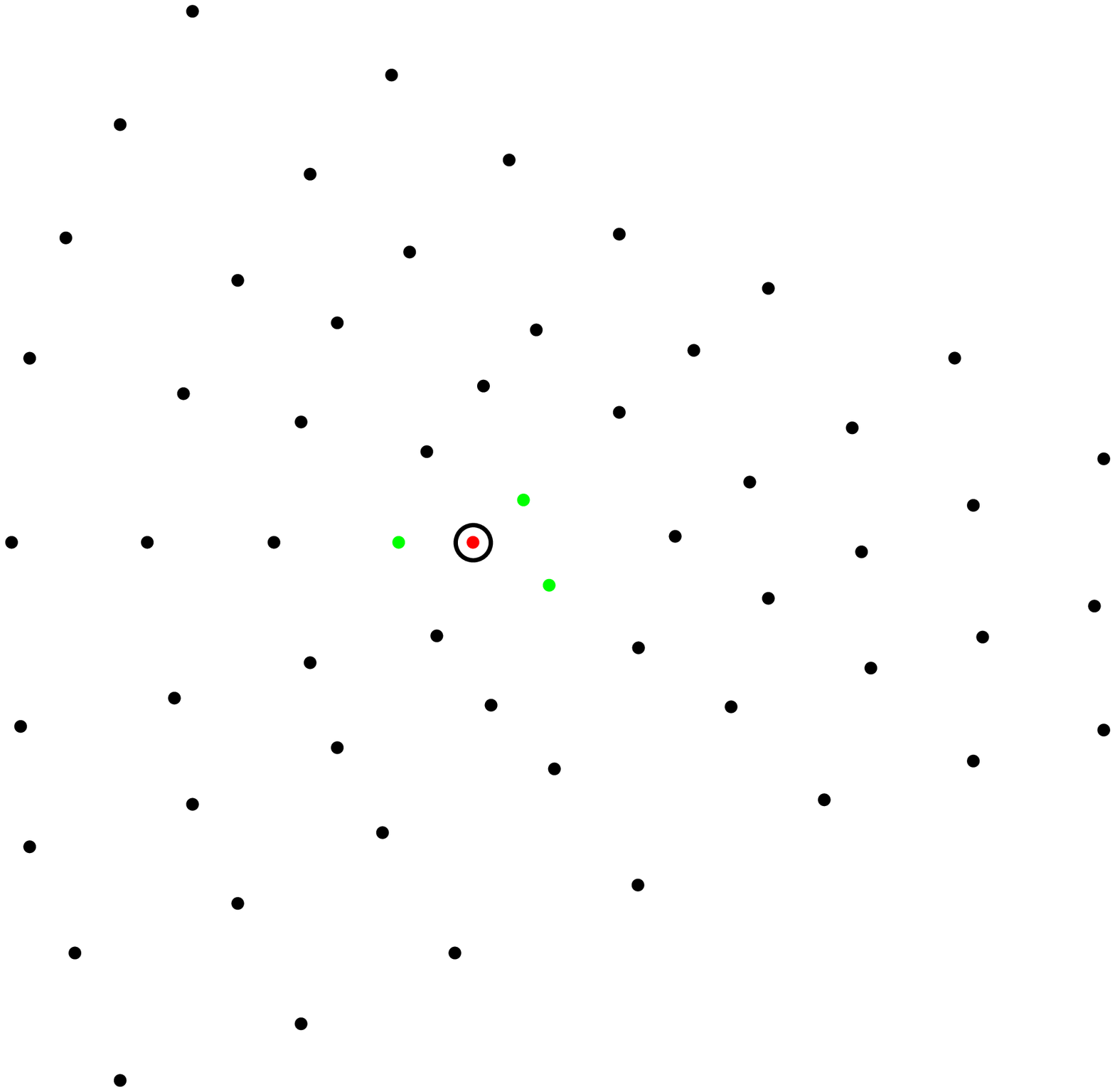}}
\hspace{1mm}
\subfigure[Insertion of a $60^\circ$ wedge into a hexagonal Bravais crystal]{\includegraphics[scale=0.18]{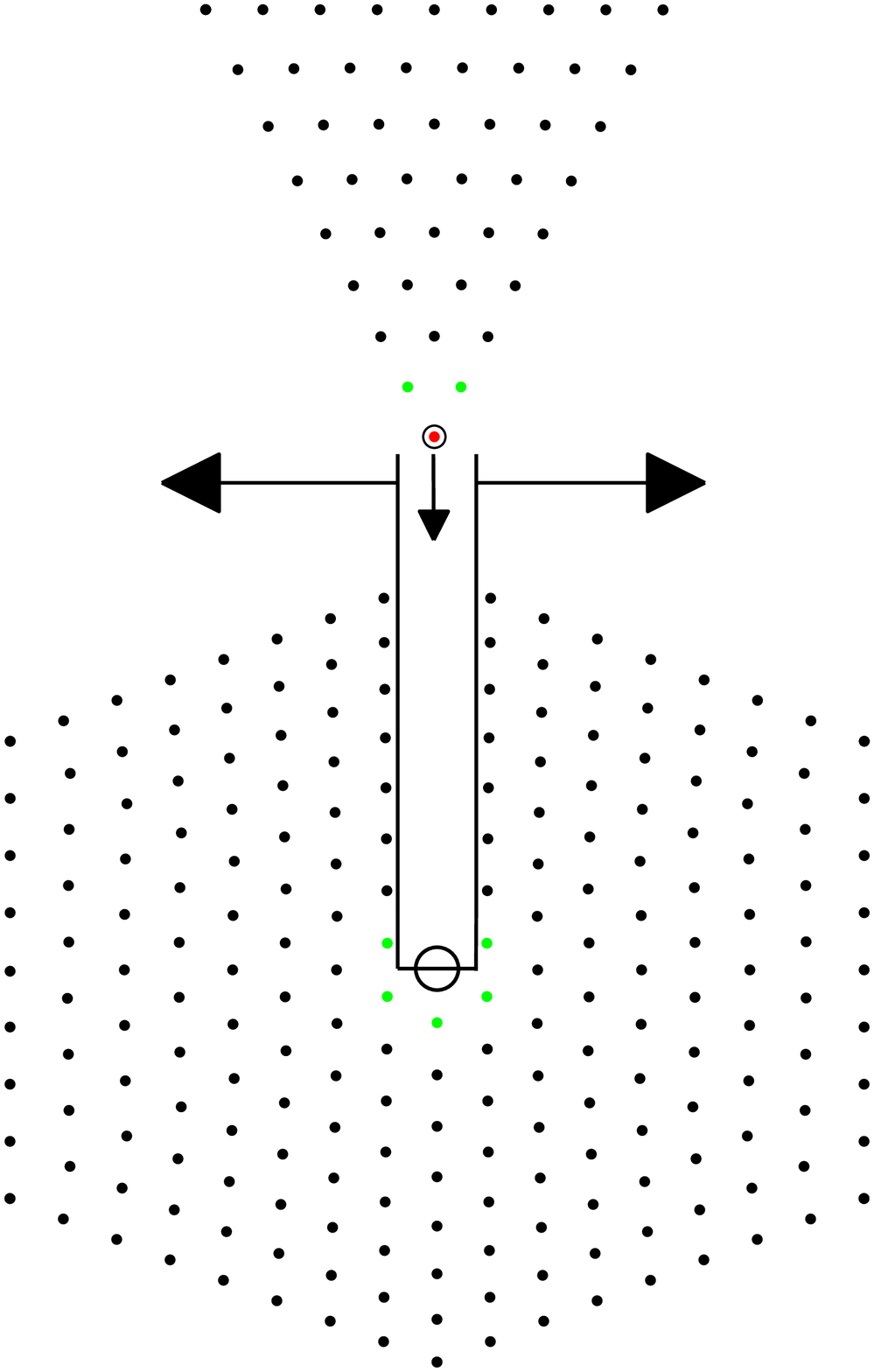}}
\hspace{1mm}
\subfigure[A +$60^\circ$ disclination]{\includegraphics[scale=0.18]{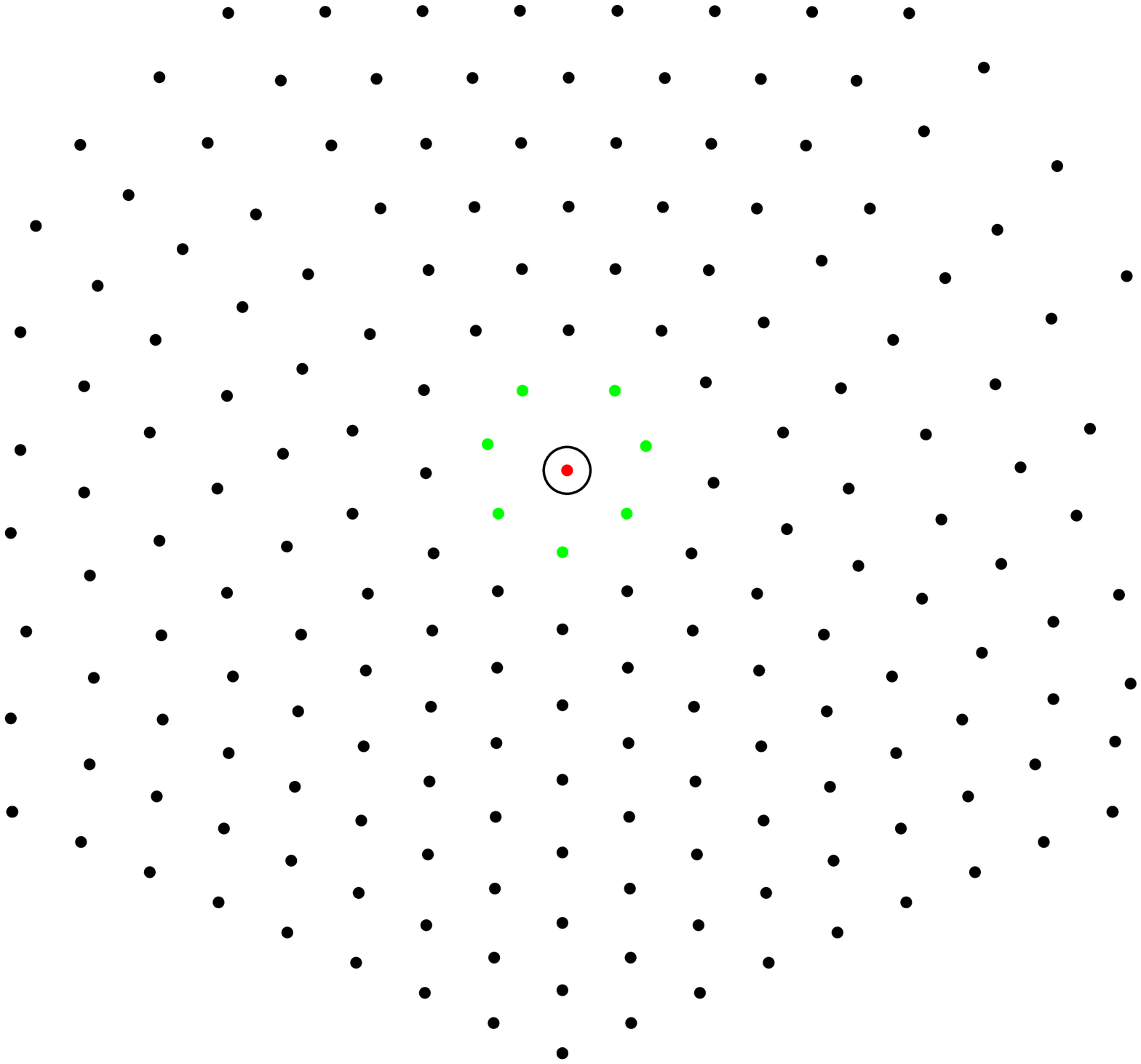}}
\caption{Formation of a negative and a positive disclination in a cubic and a hexagonal Bravais crystal, respectively.}
\label{disclicube1}
\end{figure}
Disclinations, if present, are rare in three-dimensional bulk crystalline solids. However they are ubiquitous in ordered media such as liquid crystals, magnetic materials, and two-dimensional crystals, and are as important to these solids as dislocations are to a bulk crystal.\footnote{For further reading on disclinations in a two-dimensional crystal see M. J. Bowick and L. Giomi, {\it Adv. Phy.}, 58, pp. 449-563, 2009, and W. F. Harris, in {\it Fundamental Aspects of Dislocation Theory}, J. A. Simmons et. al. (Eds.), Nat. Bur. Stand. (U.S.) Spl. Pub. 317 Vol. 1, pp. 579-592, 1970. The latter article is an original and insightful study of the application of dislocation/disclination concepts in various two-dimensional biological structures.} A disclination can be formed by inserting (or removing) a wedge in a perfect crystal, with the wedge angle equal to any of the rotational symmetry angles of the crystal. For example a $-90^\circ$ disclination in a cubic Bravais crystal is obtained, as seen in Figure \ref{disclicube1} (b), when a $90^\circ$ wedge is removed from the perfect crystal and the two lips of the cut are joined thereafter. On the other hand, a $+60^\circ$ disclination can be formed in a hexagonal crystal, as shown in Figure \ref{disclicube1} (d),  after inserting a $60^\circ$ into the perfect crystal. The line which forms the edge of the wedge is a linear defect known as disclination.

\begin{figure}[t!]
\centering
\subfigure[Negative disclination]{\includegraphics[scale=0.30]{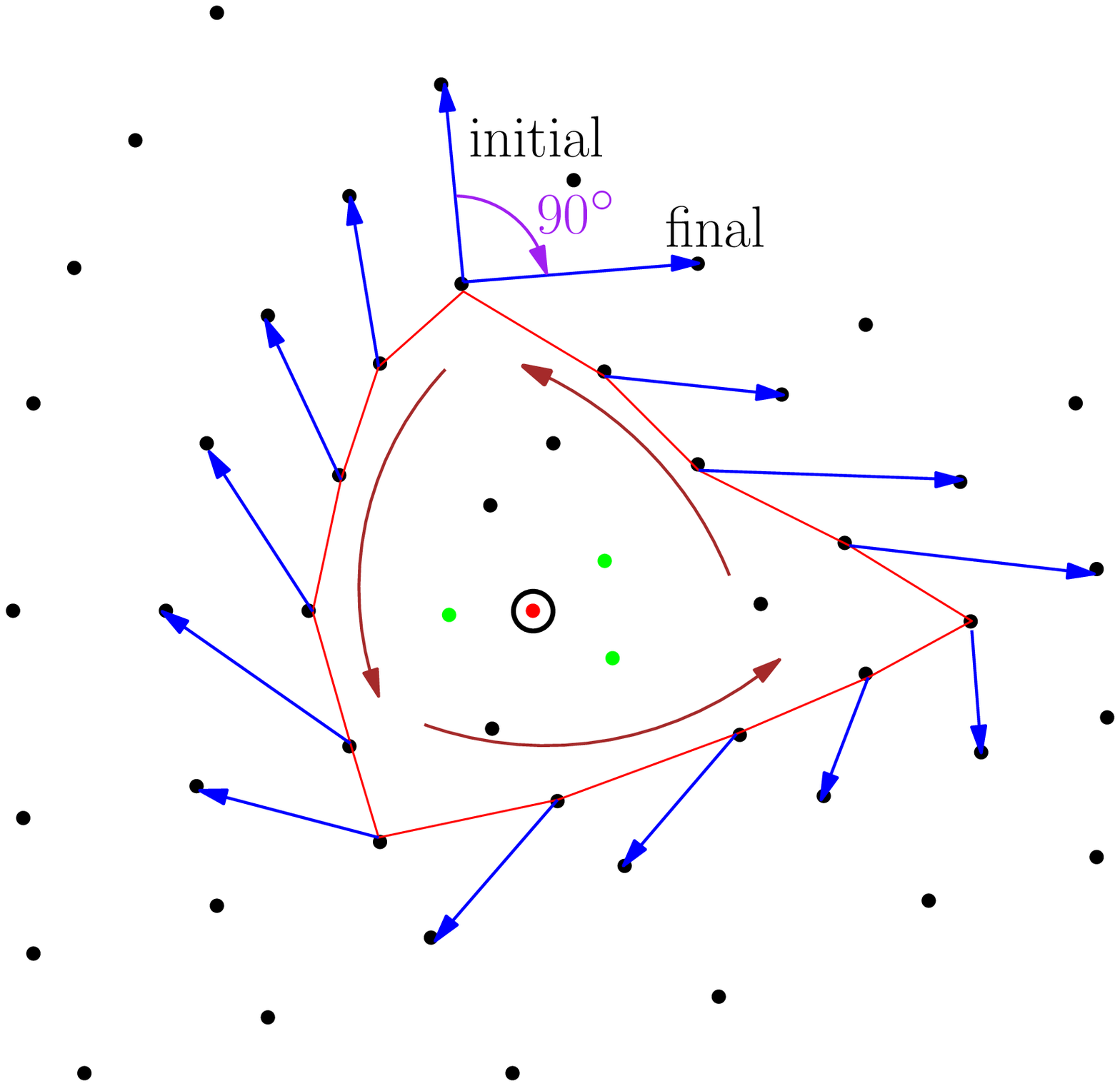}}
\hspace{15mm}
\subfigure[Positive disclination]{\includegraphics[scale=0.35]{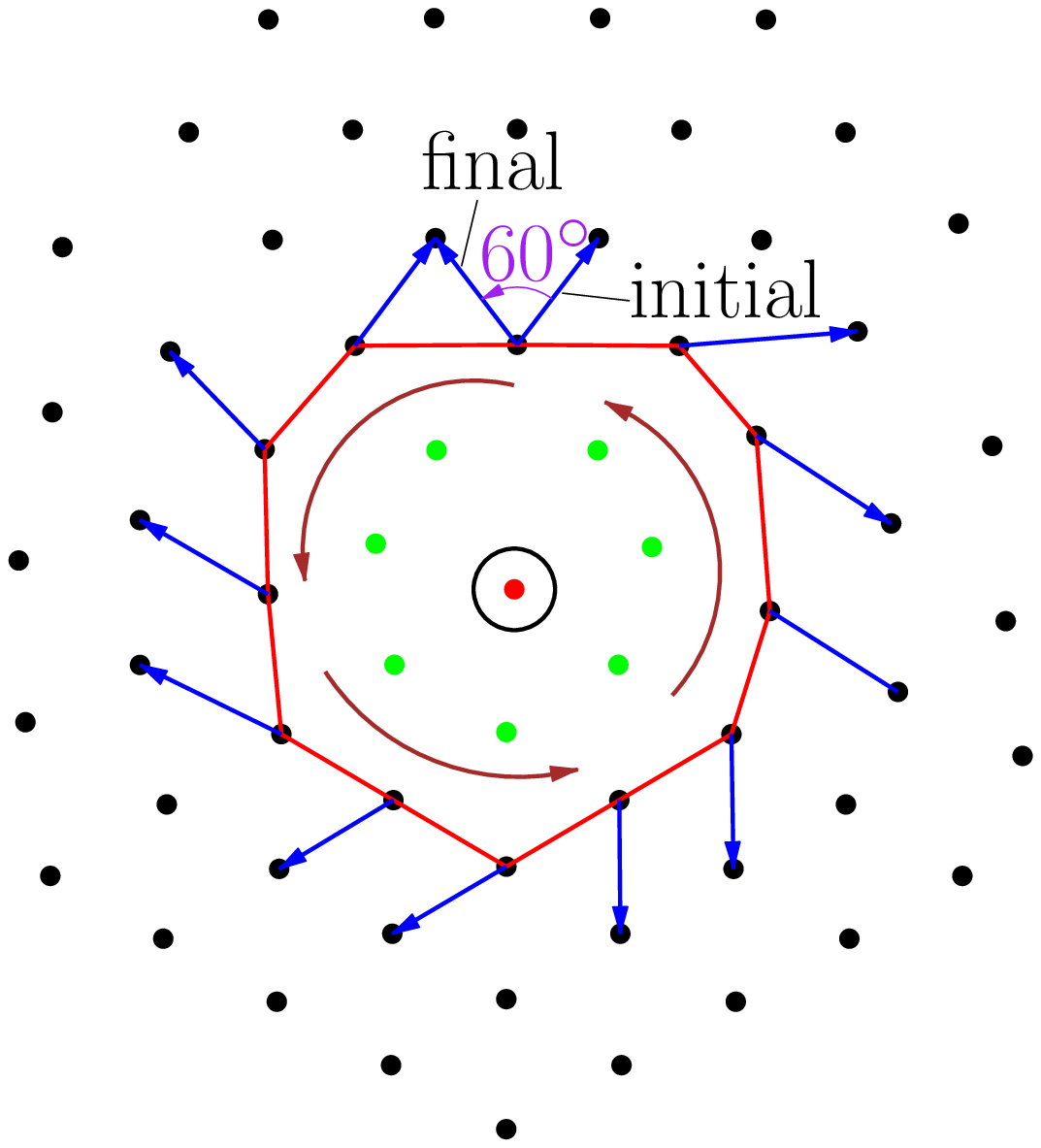}}
\caption{Parallel transport of a tangent vector along a closed loop in a disclinated crystal.}
\label{discli}
\end{figure}

Disclination is an anomaly in the rotational order of the crystal. It changes the coordination number (number of closest neighbors) of the lattice points lying on the disclination line. For instance the coordination number of the red dot in Figure \ref{disclicube1} decreases (increases) for a negative (positive) disclination. Introduction of a disclination into a perfect lattice brings about an orientational defect but preserves the translation order as well as preserve the lattice spacing. This is illustrated in Figure \ref{discli} where we observe that the parallel transportation of a tangent vector from a lattice point, along a closed curve, yields a different tangent vector. The deficiency can be characterized in terms of a rotation tensor, called the Frank's tensor. Distant parallelism is not maintained in the presence of disclination, cf. Subsection \ref{affine}. However, due to its translation preserving property, a parallelogram always closes around a disclination. The nonmetricity is again zero due to preservation of lattice spacing. As a result the affine connection associated with disclinated space should have a non-zero Riemann-Christoffel curvature but vanishing torsion and nonmetricity.

\label{discl}

\subsection{Zero-dimensional defects}

\begin{figure}[t!]
\centering
\subfigure[Vacancy defect]{\includegraphics[scale=0.5]{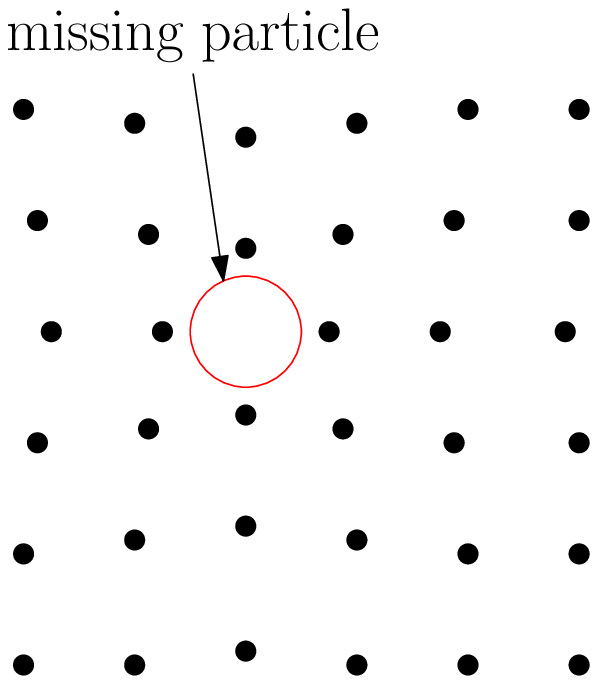}}
\hspace{8mm}
\subfigure[Interstitial defect]{\includegraphics[scale=0.5]{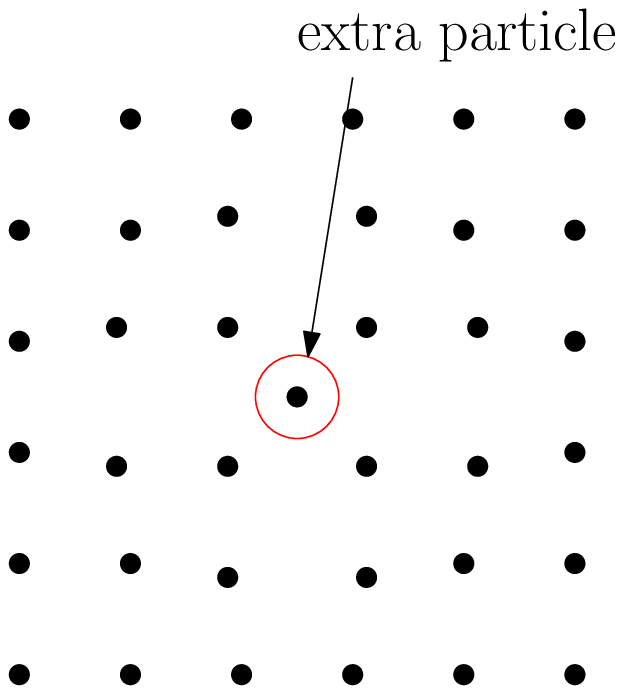}}
\hspace{8mm}
\subfigure[Substitutional defect]{\includegraphics[scale=0.5]{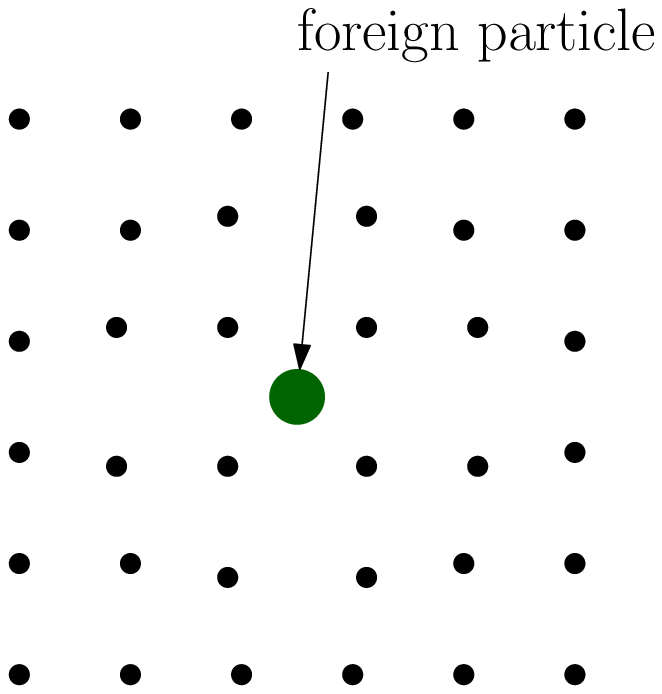}}
\caption{Zero-dimensional defects in a cubic Bravais lattice.}
\label{pdefect1}
\end{figure}

Three type of zero-dimensional defects (point defects) are shown in Figure \ref{pdefect1}. A vacancy defect is created when a particle is missing from its regular residing place in the ordered lattice arrangement. An interstitial defect results when an extra constituent atom is present in the inter-particle space of the ordered arrangement. A substitutional defect is created when a foreign particle, of a different constitution, comes to reside in the inter-particle space. All of these point defects occur frequently in crystalline solids and are significantly important during processes of thermal activation (at high temperatures) and irradiation, as well as for rate mediated phenomenon like creep. They often accumulate around higher dimensional defects (such as dislocations and grain boundaries) affecting their mobility and subsequently influencing the strength and life-time of the crystalline solid.

Introduction of point defects preserve both the translational and orientational order of the lattice, while affecting the lattice spacing. Therefore if the crystalline solid is cut to release its internal stress, and unlike dislocated/disclinated crystals mentioned above infinitely many cuts would be required presently, then different unit cells of the lattice would relax to different sizes. This is analogous to the presence of a thermal gradient within the solid.\footnote{cf. pp. 300-304 in Kr\"{o}ner, {\it op. cit.}} The affine connection related to point defects should therefore have vanishing torsion and curvature, so as to maintain translation and rotation symmetry, but non-trivial nonmetricity.

\section{Continuous distribution of defects}
The density of defects in real crystals is generally very high. For instance, the total length of dislocations in a plastically deforming crystalline solid is about $10^8-10^{10}$ mm/mm$^3$. It is therefore reasonable to talk about a continuous distribution of defects, such that the number of defects tends to infinity as the lattice spacing gets closer to zero so as to preserve the total defect content (e.g. the total Burgers vector).\footnote{This is analogous to keeping mass per unit volume fixed while reducing the lattice spacing to recover a continuum from a discrete lattice structure. For more details on the continuizing process see \S 1.4 in E. Kr\"{o}ner, {\it Modeling of Defects and Fracture Mechanics}, G. Herrmann (Ed.), Springer-Verlag, pp. 61-117, 1992.} A continuous density necessarily requires us to admit defect strengths of arbitrarily small magnitude. This requirement can impose a restriction on the possibility of continuizing the defect distribution. For instance, a density of dislocations in a crystalline solid can always be continuized because the magnitude of Burgers vector can be as small as the infinitesimal lattice spacing. On the other hand, a distribution of disclinations can never be continuized within a crystalline solid. This is due to the fact, mentioned in Subsection \ref{discl} above, that the strength of disclinations in crystals can take only finite values of rotation available from the crystal's symmetry class. Hence it does not make sense to talk about disclination density in a solid crystal. The notion of a continuous density of disclinations is valid for two-dimensional crystalline structures as well as for liquid crystals and spin structures.\footnote{cf. Bowick and Giomi, {\it op. cit.}, and  Kl\'{e}man, {\it op. cit.}} This is an appropriate place to mention a lesser known result, but significant nevertheless, that dislocation density is not a well defined field for isotropic solids. Its non-uniqueness is derived from the freedom available in imposing an arbitrary additional rotation field.\footnote{See \S 14 in W. Noll, {\it Arch. Rat. Mech. Anal.}, 27, pp. 1-32, 1967.}

We now state the central result of this article, that is the relation between defect densities and various tensor fields from differential geometry.\footnote{The connection between dislocation density and the torsion tensor from differential geometry was first pointed out by K. Kondo, {\it Proc. 2nd Japan Nat. Congr. Appl. Mech.}, pp. 41-47, 1952, and independently by B. A. Bilby et. al., {\it Proc. Roy. Soc. London, Ser. A}, 231, pp. 263-273, 1955. Notable extensions were provided by Kondo and coworkers, cf. {\it Memoirs of the Unifying study of the Basic Problems in Engineering Sciences by means of Geometry}, K. Kondo (Ed.), Vols. 1-4, Gakujutsu Bunken Fukyu-Kai, Tokyo, 1955-1968, Bilby and coworkers, cf. B. A. Bilby, {\it Prog. Sol. Mech.}, 1, pp. 331-399, 1960, Kr\"{o}ner, cf. Kr\"{o}ner, 1981, {\it op. cit.}, and Noll, {\it op. cit.}. An excellent review of the subject is given by R. de Wit, {\it Int. J. Eng. Sci.}, 19, pp. 1475-1506, 1981 and more recently by M. O. Katanaev, {\it Phys. Usp.}, 48, pp. 675-701, 2005; see also M. Kl\'{e}man, {Rev. Mod. Phy}, 80, pp. 61-115, 2008.} We begin by assuming the existence of continuous defect densities for dislocations, disclinations, and point defects, without referring to any specific solid kind. Independently, we consider a three-dimensional affine geometric space (to be called {\it material space}) defined by prescribing an affine connection (to be called {\it material connection}). The material space is assumed to be non-Riemannian with non-vanishing torsion, curvature, and nonmetricity fields. It is posited, in the light of the illustrations provided in Section \ref{lattice}, that the material space fails to be Euclidean if and only if the solid has a continuous distribution of defects. Based on our understanding of the defect character from Section \ref{lattice}, we are henceforth led to associate the torsion tensor of the material space with the dislocation density, the curvature tensor with the disclination density, and the nonmetricity tensor with the density of point defects. We will also assume existence of a metric field with which we can associate a three-dimensional Riemmanian space, which will in general be distinct from the material space. The metric is often identified with the elastic strain field in the solid, see Remark \ref{elasrem} for the related discussion. The material connection can be obtained in terms of torsion, nonmetricity, and metric field, see \eqref{friem2}. We will now consider various possibilities of defect distribution in an attempt to clarify the nature of material space.

(i) Only disclinations present ($T=0, Q=0$): With the vanishing of torsion and nonmetricity tensors the material connection is equal to the Riemmanian connection, see \eqref{friem2}; the material space is then identical to the Riemmanian space. The disclination density is given by the curvature tensor which presently has only six independent components and can be conveniently represented by a symmetric tensor with components $\theta^{pq}$ given by
\begin{equation}
\theta^{pq} = \frac{1}{4g} e^{pmj} e^{qkl} R_{mjkl}, \label{cont1}
\end{equation}
cf. \eqref{riem8} and the discussion following \eqref{riem7}. That the divergence of disclination density field is zero, cf. \eqref{riem10}, represents the fact that disclination lines cannot end arbitrarily within the solid.

(ii) Only dislocations present ($R=0, Q=0$): Unlike a Riemmanian space, the material space now has zero curvature and non-zero torsion. Vanishing of curvature implies that there exist an invertible tensor field $A^i_a$ such that the material connection is of the form \eqref{integ1}. The dislocation density field, identified with the torsion tensor (or equivalently with the tensor density $\alpha^{mi}$, see \eqref{tor3}) is given as in \eqref{tor4}. Interestingly the metric $g_{ij}$ is necessarily of the form
\begin{equation}
g_{ij} = \delta_{ab} (A^{-1})^a_i (A^{-1})^b_j, \label{cont2}
\end{equation}
where $\delta_{ab}$ is the Kronecker-delta symbol. To prove this consider a tensor with components $h_{ab} = g_{ij} A^i_a A^j_b$ and fix its value at some point in the domain as $\delta_{ab}$ (this can be always done by choosing an appropriate set of coordinate system at a point in the material space). Use \eqref{integ1} to obtain
\begin{equation}
h_{ab,k} = g_{ij;k} A^i_a A^j_b, \label{cont3}
\end{equation}
which is zero due to vanishing nonmetricity tensor. The expression \eqref{cont2} is now immediate.

(iii) Only point defects and dislocations present ($R=0$): For a vanishing curvature there exists an invertible tensor field with components $A^i_a$ such that the material connection is given by \eqref{integ1}. Assume that the metric can be written in terms of an invertible tensor $C^i_a$ as
\begin{equation}
g_{ij} = \delta_{ab} (C^{-1})^a_i (C^{-1})^b_j. \label{cont4}
\end{equation}
Substitute the metric from \eqref{cont4} and the material connection from \eqref{integ1} into \eqref{friem1} to obtain
\begin{equation}
Q_{kij} = - \delta_{ab} \left( (A^{-1})^d_i (A^{-1})^c_j + (A^{-1})^d_j (A^{-1})^c_i \right) (B^{-1})^a_c (B^{-1})^b_{d,k}, \label{cont5}
\end{equation}
where the invertible tensor $B^c_a$ is defined from
\begin{equation}
(C^{-1})^a_i = (A^{-1})^c_i (B^{-1})^a_c. \label{cont6}
\end{equation}
Differentiating \eqref{cont6} yields
\begin{equation}
L^k_{ml} = \Lambda^k_{ml}  - (A^{-1})^a_m C^k_b (B^{-1})^b_{a,l}, \label{cont7}
\end{equation}
where we have used \eqref{integ1} and $\Lambda^k_{ml} = (C^{-1})^a_{m,l} C^k_a$. Rewrite $(B^{-1})^b_{a,l} = (A^{-1})^c_l (B^{-1})^b_{a,c}$ using \eqref{integ5} to obtain the skew part of \eqref{cont7} as\footnote{cf. Eq. (27) in E. Kr\"{o}ner, {\it J. Mech. Behav. Mat.}, 5, pp. 233-246, 1994.}
\begin{equation}
T^k_{ml} = \Delta^k_{ml}  - \Upsilon ^k_{ml}, \label{cont8}
\end{equation}
where $2\Delta^k_{ml} = \Lambda^k_{ml} - \Lambda^k_{lm}$ and
\begin{equation}
2 \Upsilon ^k_{ml} = (A^{-1})^a_m (A^{-1})^c_l A^k_d [(B^{-1})^b_{a,c} - (B^{-1})^b_{c,a}] B^d_b. \label{cont9}
\end{equation}
The tensor $\Upsilon ^k_{ml}$ can be interpreted as a pseudo-torsion associated with point defects. Its relation to the nonmetricity tensor can be derived by comparing \eqref{cont8} with the skew part of \eqref{cont5}:
\begin{equation}
Q_{lmj} - Q_{mlj} = - 2 \left( g_{jk} \Upsilon ^k_{ml} + g_{mk} \Upsilon ^k_{jl} \right). \label{cont10}
\end{equation}

For the description of spherically symmetric point defects, such as vacancies and self-interstitials, it is sufficient to consider $(B^{-1})^a_c = \lambda \delta^a_c$ where $\lambda$ is a differentiable scalar field.\footnote{cf. Kr\"{o}ner, 1994, {\it op. cit.} and  M. F. Miri and N. Rivier, {\it J. Phys. A: Math. Gen.}, 35, pp. 1727-1739, 2002.}  This yields $(C^{-1})^a_i = \lambda (A^{-1})^a_i$ and $g_{ij} = \lambda^2 \delta_{ab} (A^{-1})^a_i (A^{-1})^b_j$, thereby reducing the nonmetricity tensor in \eqref{cont5} to
\begin{equation}
Q_{kij} = - 2 g_{ij} \frac{\lambda_{,k}}{\lambda} = (\ln \theta)_{,k} g_{ij}, \label{cont11}
\end{equation}
where $\theta = \lambda^{-2}$. The nonmetricity tensor obtained here is of the kind considered by Weyl (see Footnote \ref{weyl}).

\begin{rem}(Elasticity of defects) One of the most important application of the geometric theory of defects is the evaluation of internal stress field due to a distribution of defects within an elastic body. The starting point is to interpret the metric $g_{ij}$ in terms of the elastic strain tensor $\epsilon_{ij}$ as $2\epsilon_{ij} = g_{ij} - \delta_{ij}$. The material manifold is taken as the natural configuration of the body obtained by relaxing it of all its stresses. If the body is free of defects, then it can be relaxed simply by relieving itself of external forces. However, a defective body will in general be required to cut into several infinitesimal pieces to free itself of stresses. It is clear that in such a situation there is no differentiable map which connects the material manifold with the physical body in Euclidean space. Hence for a defective body, free of external influences, stress relaxation will necessarily involve the cutting operation and will lead to a strain which is incompatible.

In formal terms, we call strain $\epsilon_{ij}$ compatible if there exists a vector field $y^a$, interpreted as the motion of a material point from a fixed configuration, such that $2\epsilon_{ij} = (y^a_{,i}y^b_{,j}\delta_{ab} - \delta_{ij})$ or, according to the above interpretation of the metric, $g_{ij}=y^a_{,i}y^b_{,j}\delta_{ab}$. Incompatibility of strain, or equivalently of metric, is necessary for the emergence of residual stresses. As we have seen earlier, incompatibility of the metric is equivalent to the non-vanishing of the associated Riemannian curvature tensor $K^i_{jkl}$. The problem of finding internal stress field due to defect distribution is therefore reduced to solving the partial differential equation \eqref{friem6} for metric field where $K^i_{jkl}$ is written from \eqref{riem7} and other quantities $R$, $T$, and $Q$ are given in terms of defect densities. The stress field can be obtained from the metric, which solves the differential equation, using an assumed stress-strain constitutive law. In the absence of defects $K^i_{jkl} \equiv 0$, leading to a compatible metric field and hence to a vanishing stress field. Solving the partial differential equation has been most successful in the simplest of isotropic linear elastic solids.\footnote{cf. Kr\"{o}ner, 1981, {\it op. cit.}} Appropriate semi-inverse methods have been otherwise used frequently to obtain solutions to specific nonlinear problems.\footnote{A collection of solutions for various defect kinds can be seen in J. D. Clayton, {\it Z. Angew. Math. Mech.}, 2013, doi: 10.1002/zamm.201300142.}

\label{elasrem}
\end{rem}

\begin{rem}(Analogy with general relativity) The structure of the four-dimensional space-time in the classical general relativity theory is Riemannian, where the metric is given in terms of the tensorial gravitational potential field. Analogous to the elastic theory of defects, where defect density is the source of curvature in the material space, matter (generalized as the energy-momentum tensor) is the source of curvature in the relativistic space. There have been generalizations of the classical theory of general relativity to include torsion and nonmetricity as representations for matter spin and strain currents, respectively.\footnote{cf. F. W. Hehl et. al., {\it Rev. Mod. Phy.}, 48, pp. 393-416, 1976.}

\end{rem}

\label{cont}

\end{document}